\documentclass[twocolumn]{aastex62}

\usepackage{amsmath, natbib}

\shorttitle{Selection of AGN candidates through optical variability}
\shortauthors{S\'anchez-S\'aez et al.}

\begin{document}

\title{The QUEST--La Silla AGN Variability Survey: selection of AGN candidates through optical variability}

\author{P. S\'anchez-S\'aez}
\affiliation{Departamento de Astronom\'ia, Universidad de Chile, Casilla 36D, Santiago, Chile}
\affiliation{European Southern Observatory, Casilla 19001, Santiago 19, Chile}

\author{P. Lira}
\affiliation{Departamento de Astronom\'ia, Universidad de Chile, Casilla 36D, Santiago, Chile}

\author{R. Cartier}
\affiliation{Cerro Tololo Inter-American Observatory, National Optical Astronomy Observatory, Casilla 603, La Serena, Chile}

\author{N. Miranda}
\affiliation{Institut f\"ur Informatik, Humboldt-Universit\"at zu Berlin, 10099 Berlin, Germany}

\author{L. C. Ho}
\affiliation{Kavli Institute for Astronomy and  Astrophysics, PekingUniversity, Beijing 100871, China}
\affiliation{Department of Astronomy, School of Physics, Peking University, Beijing 100871, China}

\author{P. Ar\'evalo}
\affiliation{Instituto de F\'isica y Astronom\'ia, Facultad de Ciencias, Universidad de Valpara\'iso, Gran Bretana No. 1111, Playa Ancha, Valparaíso, Chile}

\author{F. E. Bauer}

\affiliation{Instituto de Astrof{\'{\i}}sica and Centro de Astroingenier{\'{\i}}a, Facultad de F{\'{i}}sica, Pontificia Universidad Cat{\'{o}}lica de Chile, Casilla 306, Santiago 22, Chile} 
\affiliation{Millennium Institute of Astrophysics (MAS), Nuncio Monse{\~{n}}or S{\'{o}}tero Sanz 100, Providencia, Santiago, Chile} 
\affiliation{Space Science Institute, 4750 Walnut Street, Suite 205, Boulder, Colorado 80301}

\author{P. Coppi}
\affiliation{Yale Center for Astronomy and Astrophysics, 260 Whitney Avenue, New Haven, CT 06520, USA}

\author{C. Yovaniniz}
\affiliation{Departamento de Astronom\'ia, Universidad de Chile, Casilla 36D, Santiago, Chile}

\begin{abstract}

We used data from the QUEST--La Silla Active Galactic Nuclei (AGN) variability survey to construct light curves for 208,583 sources over $\sim 70$ deg$^2$, with a a limiting magnitude $r \sim 21$. Each light curve has at least 40 epochs and a length of $\geq 200$ days. We implemented a Random Forest algorithm to classify our objects as either AGN or non--AGN according to their variability features and optical colors, excluding morphology cuts. We tested three classifiers, one that only includes variability features (RF1), one that includes variability features and also $r-i$ and $i-z$ colors (RF2), and one that includes variability features and also $g-r$, $r-i$, and $i-z$ colors (RF3). We obtained a sample of high probability candidates (hp--AGN) for each classifier, with 5,941 candidates for RF1, 5,252 candidates for RF2, and 4,482 candidates for RF3. We divided each sample according to their $g-r$ colors, defining blue ($g-r\leq 0.6$) and red sub--samples ($g-r>0.6$). We find that most of the candidates known from the literature belong to the blue sub--samples, which is not necessarily surprising given that, unlike for many literature studies, we do not cut our sample to point--like objects. This means that we can select AGN that have a significant contribution from redshifted starlight in their host galaxies. In order to test the efficiency of our technique we performed spectroscopic follow--up, confirming the AGN nature of 44 among 54 observed sources (81.5\% of efficiency). From the campaign we concluded that RF2 provides the purest sample of AGN candidates.

\end{abstract}


\section{Introduction}\label{intro}

Active galactic nuclei (AGN) are one of the most energetic phenomena in the universe and are characterized by time--variable emission in every waveband in which they have been studied. Variability studies are fundamental to understanding the extreme physical conditions of accretion disks near supermassive black holes (SMBH). Recent studies indicate that AGN variability can be well described as a stochastic process (e.g., damped harmonic oscillator or random walk; \citealt{Kelly09,Kelly14}), with characteristic time--scales ranging from days to years.

The Large Synoptic Survey Telescope (LSST; \citealt{LSST}) will revolutionize time--domain astronomy, providing for the first time the opportunity to study variable objects for a long--period of time ($\sim 10$ years), at very faint magnitudes ($r \sim 24.5$ for single images), and with a large total covered area ($>$18,000 deg$^2$). Simulations performed by the LSST AGN Science Collaboration predict detection of over $10^7$ AGN to beyond $r \sim 24$ \citep{LSSTbook}.  This is a huge increase in the number of sources available for variability analysis since current studies typically only probe to limiting magnitudes of $r \sim 21$, with total number of sources between $10$ and $10^5$ (e.g., \citealt{Cristiani97,VandenBerk04,MacLeod10,Peters15,Simm16,Caplar17,Li18}).  Given the ubiquity of variability and the large number of variable sources to be found with LSST, it is critical to characterize AGN variability and define reliable selection criteria before LSST begins operations.

Traditionally, AGN selection follows the philosophy of finding regions in UV/optical/mid--IR color--color space in which AGN can be cleanly separated from stars and galaxies (e.g., \citealt{Schmidt83,Fan99,Richards02,Richards04,Lacy04,Stern05,Smith05,Richards09,Ross12}). However, some AGN populations have observed colors that fall outside the region occupied typically by bright, ``blue'' AGN, mimicking those of stars, such as type 2 or obscured AGN, broad absorption--line quasars (BAL--QSO), high--redshift quasars (high--z QSO) \citep{Butler11,PalanqueDelabrouille11,PalanqueDelabrouille16}, and low--luminosity AGN (LLAGN), whose colors can be highly contaminated by the emission from the host galaxy. Therefore, alternative methods to identify AGN candidates missed by traditional selection techniques are required, in order to obtain complete AGN samples. One promising selection method involves the use of variability techniques.

\cite{Butler11} implemented a variability--based selection algorithm to classify high--redshift quasars in the Sloan Digital Sky Survey (SDSS; \citealt{York00}) Stripe 82 field. They used damp random walk modelling \citep{Kelly09} to separate sources showing quasar--like variability from those with temporally uncorrelated variability. Particularly, they targeted unresolved sources with redshifts in the range $2.5 \leq z \leq 3$, where color--based selection of AGN is less efficient due to stellar contamination. \cite{PalanqueDelabrouille11} used the variability structure function (e.g., \citealt{Schmidt10}) to separate quasars, variable stars, and non--variable stars, in the SDSS Stripe 82 data. They implemented a neural network algorithm that separates point--like objects by their structure function parameters. A similar technique has been used by the SDSS IV \textit{the extended Baryon Oscillation Spectroscopic Survey} (eBOSS) team to select quasar candidates with $z>2.1$ by variability \citep{Myers15,PalanqueDelabrouille16}. \cite{Peters15} used color, variability, and astrometric data from SDSS to select point--like AGN candidates. They implemented a non--parametric Bayesian Classification Kernel Density Estimation (NBC KDE), to classify 35,820 type 1 quasar candidates in the Stripe 82 field. They tested various combinations of color and variability parameters, finding that using a combination of optical colors and variability parameters improves quasar classification efficiency and completeness over the use of colors alone. More recently, \cite{Tie17} used data from the supernova fields of the Dark Energy Survey (DES; \citealt{Abbott18}) to select quasars by combining color and variability selection methods. All these previous studies have shown the potential of selecting AGN candidates through variability analyses, demonstrating that variability--based techniques can increase considerably the number of AGN candidates in the redshift range where the colors of stars are similar to those of AGN. Less clear is how deep into the low--luminosity and obscured AGN populations they can probe. This is of substantial importance given the eventual mismatch that X--ray and MIR surveys will have compared with LSST. 

In this paper we present our variability--based technique to select AGN candidates using data from the QUEST--La Silla AGN variability survey \citep{Cartier15}. In this work we aim to detect wider sets of AGN populations. Particularly, we expect to detect sources that show clear signatures of a nonstellar continuum emitting process in their centers, with emission lines broader than $\sim 1800$ km/s, regardless of their luminosity or shape in the QUEST images. We do not expect to detect many type 2 or obscured AGN candidates, since our technique requires the detection of a variable continuum component. Variability features, like the structure function, have been used to characterize the variable sources (e.g., \citealt{Cartier15,Sanchez17}). We then used a Random Forest algorithm to classify our objects as either AGN or non--AGN.  We tested three classifiers, one that includes only variability features, and two that include optical colors and variability features. The main difference of our selection technique with previous variability--based AGN selection methods is the use of light curves with higher cadence, and the exclusion of any morphology indicator. Hereby, we expect to detect more low--redshift AGN and LLAGN candidates than previous analyses. For some of our candidates we have obtained optical spectra to confirm their nature. Four of the fields observed by the QUEST--La Silla AGN variability survey correspond to the LSST Deep Drilling Fields (DDFs), whose expected cadence will be similar to the nightly cadence used by the QUEST--La Silla AGN variability survey (but extending the time baseline to 10 years)\footnote{https://www.lsst.org/scientists/survey-design/ddf}. The QUEST-La Silla AGN variability survey is an important test bed to study AGN selection in time--domain surveys, like LSST, or the Zwicky Transient Facility (ZTF, \citealt{Bellm14}), which has a depth similar to the QUEST--La Silla AGN variability survey.

The paper is organized as follows. In Section \ref{data} we describe the QUEST--La Silla AGN variability survey, and the light curve construction procedure. In Section \ref{selection} we describe the Random Forest algorithm, the variability features, and the labeled set used for the selection. We also discuss the performance of our Random Forest classifiers, and we comment about the selected candidates. In Section \ref{cand_conf} we provide the results on confirming the nature of some of our candidates by using public data and spectroscopic follow--up. In Section \ref{comparison}, we provide a comparison of our results with previous works. Finally, in Section \ref{discussion} we summarize the main results.

\section{Data}\label{data}

\subsection{The QUEST--La Silla AGN variability survey}\label{quest}

Between 2010 and 2015 we carried out ``The QUEST--La Silla AGN variability survey'' (hereafter QUEST--La Silla), using the wide--field QUEST camera mounted on the 1m ESO-Schmidt telescope at La Silla Observatory \citep{Cartier15,Cartier16}. The survey used a broadband filter, the $Q$ band, similar to the union of the $g$ and $r$ SDSS filters. Our survey includes the COSMOS, ECDF--S, ELAIS--S1, XMM--LSS and Stripe 82 fields. These are some the most intensively observed regions in the southern sky. Our QUEST fields are much larger than the nominal fields, but we will still adopt the same names, with a surveyed area of $\sim 14$ deg$^2$ per field, with the exception of the XMM--LSS field, which covers an area of $\sim 38$ deg$^2$. One of the advantages of our survey over other surveys is the intense monitoring used, observing the fields every possible night (but see the binning strategy described below). Individual images reached a limiting magnitude between $r \sim 20.5$ and $r \sim 21.5$ mag for an exposure time of 60 seconds or 180 seconds, respectively. 

The aims of our survey are: 1) to test and improve variability
selection methods of AGN, and find AGN populations missed by other
optical selection techniques \citep{Schmidt10,Butler11,PalanqueDelabrouille11}; 2) to obtain a large number of well--sampled light curves, covering time--scales ranging from days
to years; 3) to study the link between the variability properties
(e.g., characteristic time--scales and amplitudes of variation) with
physical parameters of the system (e.g., black-hole mass, luminosity,
and Eddington ratio). 

\cite{Cartier15} presented
the technical description of the survey, the full characterisation of
the QUEST camera, and a study of the relation of variability with
multi-wavelength properties of X--ray selected AGN in the COSMOS
field. In \cite{Sanchez-Saez18} we performed a statistical analysis of the connection between AGN variability and physical properties of the central SMBH, where we found that the amplitude of variability at one year time--scale (A) depends primarily on the rest--frame emission wavelength ($\lambda_{rest}$) and the Eddington ratio, where A anticorrelates with both $\lambda_{rest}$ and $L/L_{\text{Edd}}$. 

\subsection{Light curve construction}\label{lc}

We reduced the data from the QUEST--La Silla using our own customized pipeline, following the same procedure described by \cite{Cartier15}, which includes dark subtraction, flat--fielding, and astrometric and photometric calibration. To calibrate the photometry, we used public photometric SDSS catalogs \citep{Gunn98,Doi10} for the COSMOS, Stripe 82 and XMM--LSS fields, and public catalogs from the first year of DES \citep{Abbott18} for the ELAIS--S1 and ECDF--S fields. We performed aperture photometry using SExtractor \citep{sextractor}, with the same optimal aperture found by \cite{Cartier15} for the QUEST camera ($\sim 6\arcsec.18$). We then constructed light curves for all the sources from the SDSS and DES catalogs with detections in the QUEST--La Silla data, using the same methodology as in \cite{Cartier15}. In summary, we constructed light curves by cross--matching the SDSS and DES catalogs with every QUEST--La Silla catalog, that we generated for each observation, for which we knew their associated Julian dates, using a radius of 1\arcsec. We then constructed light curves for each source, keeping only those epochs where the SExtractor FLAG parameter was equal to zero, to prevent false detection of variability due to bad photometry. Finally, we only saved those light curves with more than three epochs. From the SDSS catalog, we could obtain single--epoch photometry of every source in the COSMOS, XMM--LSS, and Stripe 82 fields in the $u$, $g$, $r$, $i$, and $z$ bands, and from the DES catalog we obtained single--epoch photometry in the $g$, $r$, $i$, and $z$ bands for the ELAIS--S1 and ECDF--S fields.

We decided to bin our light curves using three--day bins, in order to reduce the noise in our light curves produced by several factors, including changes in atmospheric conditions and the relatively low cosmetic quality of the QUEST CCD camera chips. This might affect the detection of variability of sources with short time--scale variations, like some variable stars (e.g., RR Lyrae or Cepheid stars), however, we do not expect to detect many variable stars in the QUEST--La Silla fields (e.g., \citealt{Medina18}). Moreover, in this work we are focused in the detection of sources with long time--scale variations (with time--scales of months or years), thus the three--day binning does not affect our detection of AGN. 

In this work, we excluded the Stripe 82 field, since it is a crowded field, and requires point spread function (PSF) photometry. We generated a total of $277,629$ light curves for sources located in the COSMOS, ECDF--S, ELAIS--S1, and XMM--LSS fields. In order to have statistically significant variability features of the sources, we decided to include in our analysis only those light curves with at least 40 epochs and a length greater than or equal to 200 days, after the three days binning was applied (hereafter ``well--sampled'' light curves). There are $208,583$ well--sampled light curves in the four fields. The median, mean and standard deviation of the number of epochs of each light curve are 118, 119.3, and 47.2, respectively; and the median, mean and standard deviation of the total length of each light curve are 1283.7, 1306.4, and 254.3, respectively. In Table \ref{tab:num_lc} we summarize the total number of light curves and the number of well--sampled light curves in each field.

\begin{table}
\caption{Number of light curves per field.} \label{tab:num_lc} 
\begin{center}

\begin{tabular}{ccc} \hline
\hline

Field & total light curves &  well--sampled light curves  \\
\hline

COSMOS & 68,514 & 45,323 \\
XMM--LSS & 104,962 & 82,697 \\
Elais--S1 & 49,504 & 38,106 \\
ECDF--S & 54,649 & 42,457 \\

\hline

Total & 277,629 & 208,583 \\

 \hline
 \hline

\end{tabular}


\end{center}

\end{table}

\section{Selection of AGN candidates}\label{selection}

We implemented a supervised automatic classification using a Random Forest algorithm (RF; \citealt{randomforest}) to classify our 208,583 objects with well--sampled light curves as either AGN or non--AGN according to their variability features and optical colors. We did not include a morphological parameter during the classification (e.g., SExtractor CLASS\_STAR parameter), in order to be able to detect sources with AGN--like variability with extended shapes. We tested three classifiers: one that includes only variability features, and two that include optical colors and variability features. In the following sub--sections we describe the selection methodology, the features used in our analysis, and the results of the classification for sources from the QUEST--La Silla survey.

\subsection{Random Forests}\label{RF}

A decision tree is a hierarchical structure that performs successive partitions on the data, each of them according to a certain criteria, such as a cut--off value in one of the descriptors or features. In this way, the data are divided into smaller and smaller subsets as the tree goes deeper, until it reaches the leaves of the tree. Each of the leaves is associated with a single class. A given class, however, may be associated with several leaves. Thus the elements that fall on any of the leaves corresponding to a particular class will be classified as belonging to that class.

A RF algorithm consist of a collection of single decision trees, where each tree is trained using a random sub--set of sources, sampled with repetition, from a training set (a set of objects with known classification, selected from a labeled set), and a random selection of features. The final classification function of the algorithm weighs each of these results according to the size of the sub--set used by each tree, and generates an average score, which can then be interpreted as the probability that the input element belongs to a certain class (predicted class probability, $P_{\text{RF}}$). Then, the classifier is validated using a sub--set of the labeled set that was not used to train (the test set). Finally, a prediction is made on the unlabeled data. RF has several advantages, it can handle thousands of features, it provides a ranking of feature importance during the classification, it does not need to scale the feature values to the same ``units'', it handles numerous objects, and it is easily parallelizable.

For the selection of AGN candidates we used the \textit{scikit-learn}\footnote{http://scikit-learn.org/stable/modules/generated/sklearn. ensemble.RandomForestClassifier.html} Python package implementation of RF.  We performed a hyperparameter selection procedure in order to obtain the optimal values for the RF classifier, by means of a K--Fold Cross--Validation procedure\footnote{https://scikit-learn.org/stable/modules/cross\_validation.html} (with $k=5$ folds) and using the ``accuracy'' (see its definition in Section \ref{rf_perform}) as the target score to optimize. This hyperparameter selection procedure was executed as part of the model training phase (i.e. on the training set). In this procedure, the training set is divided in $k$ folds, using $k-1$ of them to compute the RF model, and testing it in the remaining data (the validation set). This is done $k$ times, using every time a different fold as validation set. The parameters considered in this cross--validated search include the number of trees in the forest, and the number of features to consider when looking for the best split in a tree. In order to take into account the class imbalance in the classification process, we initialized the class weight hyperparameter as ``\textit{balanced\_subsample}''.

The variability features used by the RF classifier are described in the following section (\ref{var_feat}), and are listed in Table \ref{tab:features}. We trained the RF classifier using a labeled set of type 1 AGN and stars with spectroscopic classification from SDSS and with well--sampled light curves from the QUEST--La Silla survey (see Section \ref{lab_set}). During the RF classifier training, we used $70\%$ of the labeled set as a training set, and then we tested the performance of the classifier using the remaining $30\%$ of the labeled set (the test set), as normally done during supervised learning procedures. We then applied the trained RF classifier to our unlabeled set, composed by our 208,583 sources with well--sampled QUEST--La Silla light curves, to classify them as  either AGN or non--AGN. As a result, we obtain a predicted class and the predicted class probability ($P_{\text{RF}}$) associated to each source of the unlabeled set.

\subsection{Variability features}\label{var_feat}

In order to have a complete description of the variability of our sources, we used several variability features. Following the same approach of \cite{Sanchez17} and \cite{Sanchez-Saez18}, we used two parameters related to the amplitude of the variability, $P_{var}$ and the excess variance ($\sigma_{\text{rms}}$), and one parameter that describes the shape of the variability between two observations separated by a given time, the structure function (SF).

In particular, $P_{var}$ (see \citealt{Sanchez17} and references therein) corresponds to the probability that the source is intrinsically variable; it considers the $\chi^2$ of the light curve, and calculates the probability $P_{var}=P(\chi^2)$ that a $\chi^2$ lower or equal to the observed value could occur by chance for an intrinsically non--variable source. 

$\sigma_{\text{rms}}$ is a measure of the intrinsic variability amplitude (see \citealt{Sanchez17} and references therein), and it is calculated as $\sigma^2_{rms}=(\sigma_{LC}^2-\overline{\sigma}_{m}^2)/\overline{m}^2$, where $\sigma_{LC}$ is the standard deviation of the light curve, $\overline{\sigma}_{m}$ is the mean photometric error, and $\overline{m}$ is the mean magnitude. 

The SF (e.g., \citealt{Schmidt10}) is the average variability amplitude between two observations separated by a given time ($\tau$), and it can be modelled as a power--law: $\text{SF}(\tau)=A_{\text{SF}}\left( \frac{\tau}{1\text{yr}}\right)^{\gamma_{\text{SF}}}$, where $A_{\text{SF}}$ corresponds to the amplitude of the variability at 1 year time--scale, and $\gamma_{\text{SF}}$  is the logarithmic gradient of this change in magnitude.

We also used some variability features from the Feature Analysis for Time Series (FATS; \citealt{fats}) Python package, related with the amplitude of the variability (e.g., the mean variance and the percent amplitude) and the structure of the light curve (e.g., the linear trend and the auto--correlation function length), as well as the period of the Lomb--Scargle periodogram \citep{VanderPlas18}, derived by using the AstroML module for Python \citep{astroML}. A list of all the variability features used in this work is shown in Table \ref{tab:features}, together with a brief description of each feature and its reference.

\begin{table*}
\caption{List of features.} \label{tab:features} 
\begin{center}

\begin{tabular}{lll} \hline
\hline

Feature & Description & Reference \\
\hline

$P_{var}$ & Probability that the source is intrinsically variable & \cite{McLaughlin96} \\
$\sigma_{\text{rms}}$  & Measure of the intrinsic variability amplitude. &  \cite{Allevato13} \\
$A_\text{SF}$ & Amplitude of the variability at 1 year, derived from the SF  & \cite{Schmidt10} \\
$\gamma_\text{SF}$ & Logarithmic gradient of the change in magnitude, derived from the SF & \cite{Schmidt10} \\
Std* & Standard deviation of the light curve ($\sigma_{LC}$) & \cite{fats} \\
Meanvariance* & Ratio of the standard deviation to the mean magnitude ($\sigma_{LC} / \overline{m}$)   & \cite{fats} \\
MedianBRP* & Fraction of photometric points within amplitude/10 of the median magnitude & \cite{Richards11}\\
Autocor--length* &  Lag value where the auto--correlation becomes smaller than $e^{-1}$ & \cite{Kim11} \\
StetsonK* & A robust kurtosis measure & \cite{Kim11}  \\
$\eta^e$* & Ratio of the mean of the square of successive differences to the variance of data points & \cite{Kim14} \\
PercentAmp* & Largest percentage difference between either the max or min magnitude and the median & \cite{Richards11} \\
Con* & number of three consecutive data points that are brighter or fainter than 2$\sigma_{LC}$ &  \cite{Kim11}\\
LinearTrend* & Slope of a linear fit to the light curve & \cite{Richards11} \\
Beyond1Std* & Percentage of points beyond one $\sigma_{LC}$ from the mean & \cite{Richards11} \\
Q31* & Difference between the third quartile and the first quartile of a light curve & \cite{Kim14} \\
PeriodLS & Period from the Lomb--Scargle periodogram & \cite{VanderPlas18} \\

 \hline
 \hline

\end{tabular}


\end{center}
\textbf{Note.} (*) Features from FATS 

\end{table*}

\subsection{Labeled set}\label{lab_set}

To train our RF classifier we need a labeled set, which has to be representative of the populations that we want to classify. Since in this analysis we only include extragalactic fields, we do not expect to detect a high fraction of variable stars, because of their low density at high Galactic latitudes (e.g., RR Lyrae or Cepheid stars, see \citealt{Medina18} and references therein). Moreover, since we implemented a three--day binning to our light curves, the detection of variable signals with short time--scales is not possible. Therefore, any variable star with a short period will have a light curve that will not be very different from a non--variable star. Only variable stars with long periods would be detectable using our QUEST--La Silla light curves. We cross--matched the positions of the 208,583 sources with well--sampled light curves with the General catalogue of variable stars (Version GCVS 5.1, \citealt{Samus17}), which provides a detailed compilation of catalogs of variable stars in the Galaxy. We found that only three known variable stars are present in our data, one RR Lyrae and two cataclysmic variables. Therefore, we did not include variable stars in our RF classifiers.

In this analysis, galaxies are not included in the labeled set, since in general their variability and color properties will be similar to those of stars, unless they host an AGN (which might not have been previously detected). Therefore, we constructed a labeled set composed by stars and type 1 AGN (i.e. AGN with broad permitted emission lines). We decided to include only type 1 AGN since we want to characterize properly the variability of the optical continuum emission, which cannot be detected in most type 2 or obscured AGN. 

\begin{figure}
\begin{center}
\includegraphics[scale=0.4]{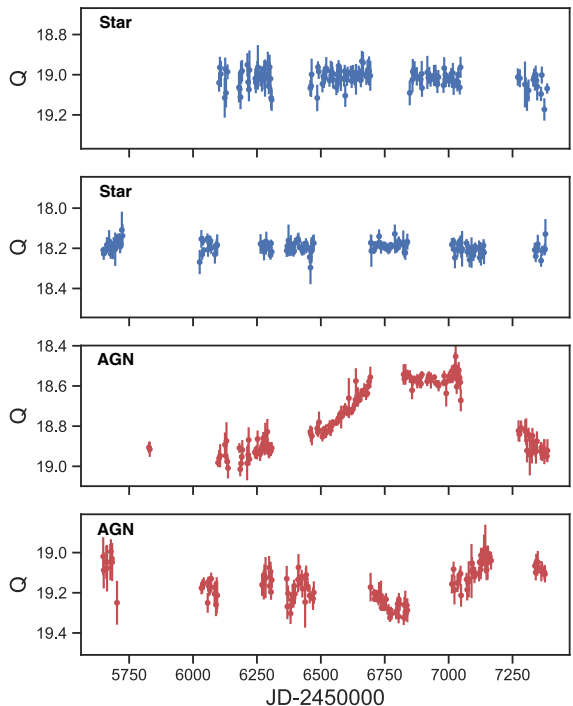}
\caption{Example of four light curves from the QUEST--La Silla labeled set: two stars (blue dots, top panels) and two AGN (red dots, bottom panels). \label{figure:train_lc}}
\end{center}
\end{figure}

Three of our fields (COSMOS, Stripe 82, and XMM--LSS) have spectroscopic information from SDSS. We constructed light curves for sources with spectral classification from the SDSS--DR14 database \citep{sdss-dr14}. There are 3,313 type 1 AGN and 3,332 stars with at least three epochs in the QUEST--La Silla light curves, and 2,405 type 1 AGN and 2,608 stars with well--sampled light curves. We considered the sources with well--sampled light curves to define a labeled set for the RF classifier training. As mentioned in Section \ref{RF}, 30\% of the labeled set was used as a test set and 70\% as training set for the RF modelling. Figure \ref{figure:train_lc} provides examples of QUEST--La Silla light curves for four sources of the labeled set. 

It is well--known that a fraction of AGN are misclassified in the SDSS databases, therefore we cross--match our labeled sample with the Million Quasars Catalog (MILLIQUAS v5.7 update\footnote{https://heasarc.gsfc.nasa.gov/W3Browse/galaxy\-catalog/milliquas.html}, 7 January 2019, \citealt{Flesch15}, see Section \ref{other_cat} for further details), in order to estimate the fraction of AGN in the labeled set that are not, in fact, AGN.  There are 57 AGN in the labeled set, with well--sampled light curves, that are not present in MILLIQUAS, which correspond to the 2.4\% of the AGN in the labeled set. 21 of these 57 sources are classified as variable according to their variability features, thus we can estimate that less than 2\% of the AGN in the labeled set are misclassified as AGN.

\begin{figure*}
\begin{center}
\includegraphics[scale=0.5]{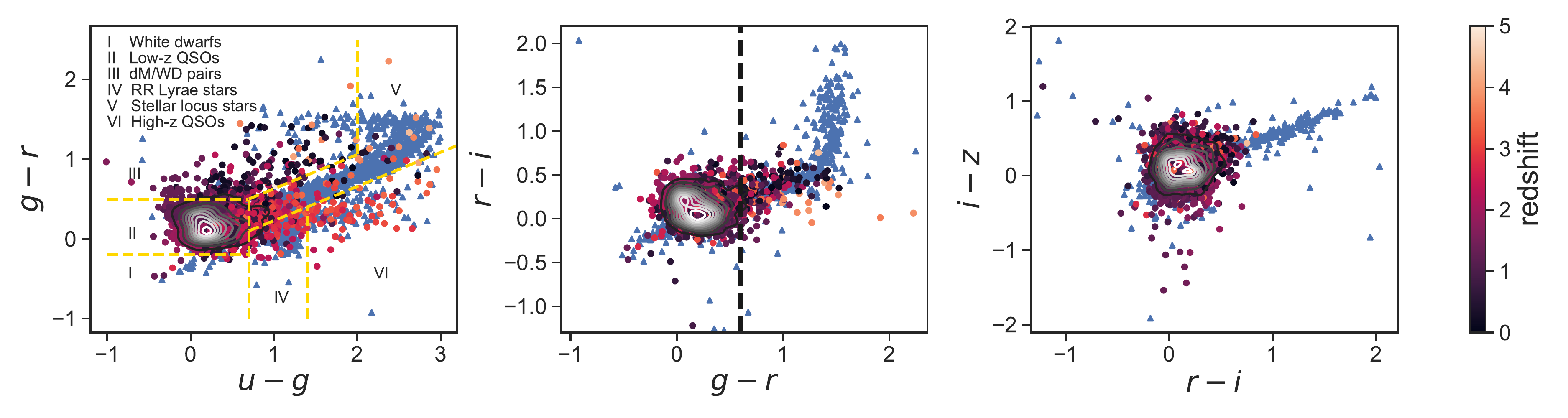}
\caption{Color--color diagrams of the labeled set. In the left panel we show $u-g$ versus $g-r$, in the middle panel $g-r$ vs $r-i$, and in the right panel  $r-i$ vs $i-z$. The stars are represented by blue triangles, and the AGN are represented by circles whose colors depend on the redshift of each source. The contour plots show the distribution of AGN. In the left panel we show  with yellow dashed lines the division used in \cite{Sesar07} to identify regions of the $u-g$ versus $g-r$ diagram dominated by a particular type of source. In the middle panel, the black dashed line shows the position where $g-r=0.6$.  \label{figure:train_color}}
\end{center}
\end{figure*}

In Figure \ref{figure:train_color} we show three color--color diagrams
of the labeled set: $u-g$ vs $g-r$, $g-r$ vs $r-i$, and $r-i$ vs
$i-z$. As a reference, we mark the regions of the $u-g$ versus $g-r$
color--color diagram dominated by a particular type of source, from
\cite{Sesar07}. We can see that a high fraction of the AGN in the
labeled set are located in a region of the $u-g$ vs $g-r$ diagram
where low--redshift, luminous AGN (II) are the dominant population,
which corresponds to 78.8\% of the AGN in the labeled set. This
correponds to the classical color--color selection of ``blue'' AGN.
Moreover, we can see in the figure that several high--redshift
($z_{spec}>2.5$) AGN in the labeled sample are located in the region
where high--redshift luminous AGN are the dominant component (VI), as
expected, but a non negligible fraction is located in other regions,
where binary stars or cool dwarf stars (III), RR Lyrae stars, and main sequence stars or
``stellar locus'' (V) are the dominant population. 

On the other hand, we can see in the right panel of Figure
\ref{figure:train_color} that most of the AGN in the labeled set are
located in a region where $r-i\lesssim0.7$ and $i-z\lesssim0.8$,
cleanly isolating a sub--population of cool dwarf stars. This can be
understood considering that stellar colors become monotonically redder
as the effective temperature decreases \citep{Covey07}, thus we
normally observe a high concentration of cool dwarf stars in a region
around $g-r\sim 1.5$, with $r-i \gtrsim 0.8$. Besides, extragalactic
sources (i.e., galaxies and AGN) are normally located in regions of
the color--color space with $r-i \lesssim 1.0$ (e.g.,
\citealt{Rahman16}), since their integrated emission typically has a
low contribution from cool dwarf stars. Therefore, we can use $r-i$
and $r-z$ colors to separate AGN from cool dwarf stars. The separation
of AGN and stars from the general ``stellar locus'' is more complicated
if we only use optical colors, since there is a high overlap between
these two populations in the different color--color diagrams,
particularly in the $u-g$ vs $g-r$ diagram, as already
discussed. Thus, including variability information in the selection of
AGN candidates will be extremely useful to improve AGN selection.

\subsection{Performance of the Random Forest classifier}\label{rf_perform}

We tested three different RF classifiers. The first one includes only variability features (hereafter RF1). The second one includes variability features and the $r-i$ and $i-z$ colors (hereafter RF2). And the third classifier includes variability features and the $g-r$, $r-i$, and $i-z$ optical colors (hereafter RF3).  We exclude $u-g$ since we do not have photometry in the $u$ band for the Elais--S1 and ECDF--S fields, and because our labeled set does not cover properly the $u-g$ space, compared to the unlabeled set, thus including it might produce poor results. For this reason we did not test a pure color selection, as $u$ band is highly discriminating for selecting AGN (e.g., \citealt{Richards02,Richards09,Ross12}).

As can be seen, the difference between RF2 and RF3 is the exclusion of the color $g-r$ in RF2. As mentioned in the previous section, the $r-i$ and $i-z$ colors can easily separate cool dwarf stars and AGN. However, the separation of AGN and stars from the stellar locus is difficult when we use optical colors, particularly for the case of $u-g$ and $g-r$. Thus, with RF2 we can  test whether avoiding the use of $g-r$ can improve the detection of redder AGN populations. On the other hand, RF1 excludes optical colors, thus the amount of information used by this classifier is lower compared to RF2 and RF3. Optical colors have been exhaustively used in the literature for the selection of AGN candidates (e.g., \citealt{Fan99,Richards02,Richards04,Smith05,Richards09,Kirkpatrick11,Bovy11,Ross12}), thus with RF1 we can test whether single--band variability--based techniques can provide results as competitive as the ones obtained using optical colors.

As mentioned in Section \ref{RF}, we trained each classifier using $70\%$ of the labeled set as a training set, and the remaining $30\%$ of the labeled set as a test set. The selection of the training and test sets is done randomly, using the  ``train\_test\_split'' procedure of \textit{scikit--learn}. The labeled set, by definition, has the same limiting magnitude of the QUEST--La Silla images ($r \sim 21$), therefore the training and test sets have limiting magnitudes of $r \sim 21$ .

Since we are interested in selecting only AGN candidates, for the rest of the analysis we will refer to stars, and any source that is not an AGN as non--AGN.

\subsubsection{RF1: selection of AGN based solely on variability}

\begin{figure}
\begin{center}
\includegraphics[scale=0.65]{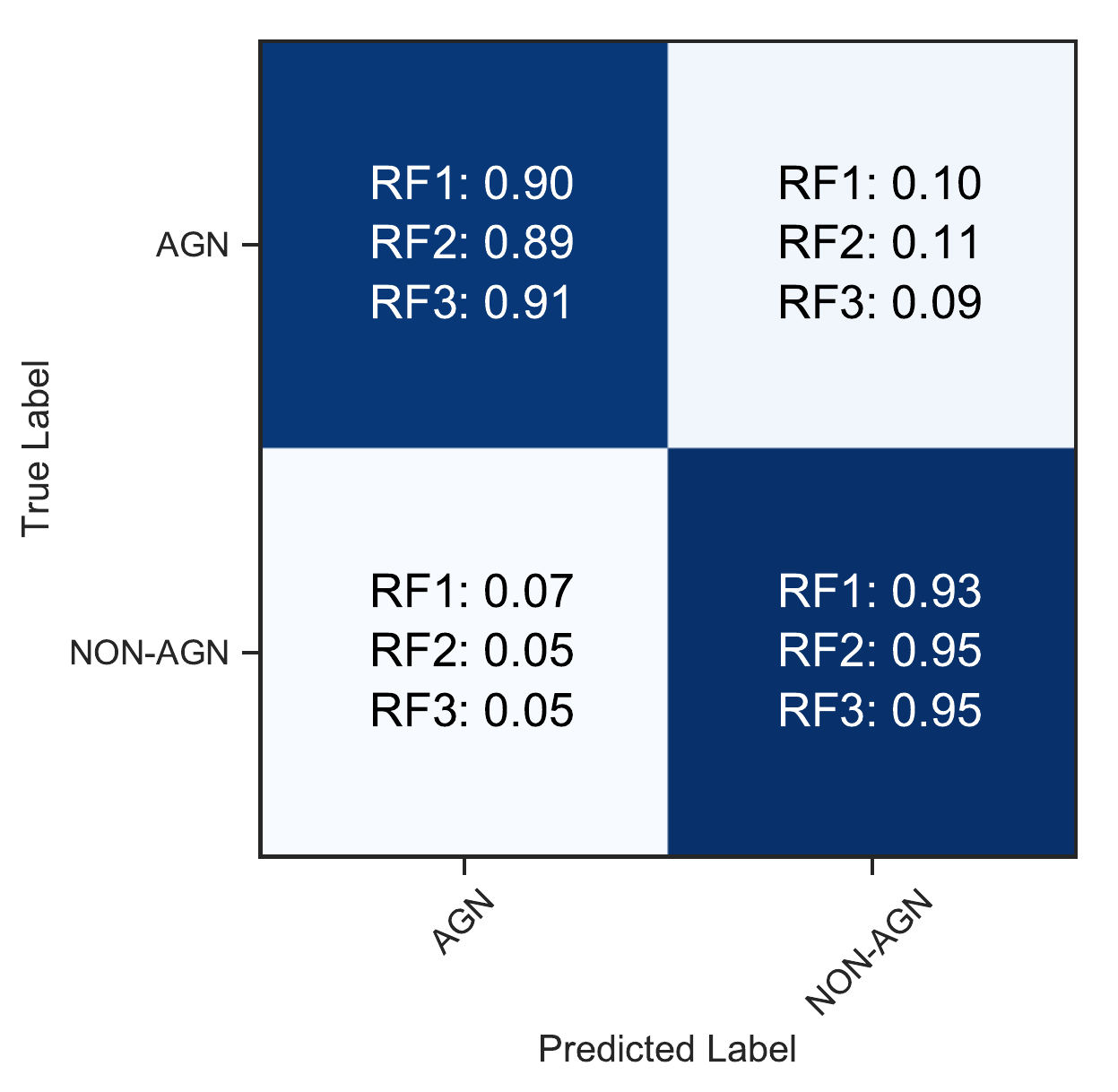}
\caption{Confusion matrix from testing the RF1, RF2, and RF3 in the test set. True Label represent the classification done from SDSS spectra, and Predicted Label is the outcome of each classifier. \label{figure:conf_matrix_all}}
\end{center}
\end{figure}

Our first RF classifier (RF1) includes only variability features. We show the results from this classifier using a confusion matrix, which is shown in Figure \ref{figure:conf_matrix_all} (see its RF1 results). It can be seen that AGN (true positives) are in general well classified, and also that the fraction of non--AGN classified as AGN (false positives) is very low.

We also computed the following scores to assess our classifiers: accuracy ($A$), precision ($P$), recall ($R$), and F1. These scores are defined by means of the True Positives (TPs: known AGN classified as AGN by the RF classifier), the False Positives (FPs: known non--AGN classified as AGN), the True Negatives (TNs: known non--AGN classified as non--AGN), and the False Negatives (FNs: known AGN classified as non--AGN):

\begin{equation}
\begin{aligned}
A&=\frac{TPs+TNs}{Total \: Sample} \\
P&=\frac{TPs}{TPs+FPs} \\
R&= \frac{TPs}{TPs+FNs}\\
\text{F1}&=2\times \frac{P\times R}{P+R} 
\end{aligned}
\label{eq:scores}
\end{equation}

Table \ref{tab:scores} shows the computed scores for the RF1 classifier. From these scores, and from the confusion matrix, we can say that RF1 presents a low fraction of False Positives, thus, the sample of predicted AGN has low contamination from non--AGN. However, we tend to miss a fraction of real AGN ($\sim 10\%$). This results from the difficulty of detecting a variable signal from AGN with low amplitude variability, and since we are only considering variability properties for the classification, they could be classified as non--AGN.

It is important to consider that we are testing the RF1 classifier in a sample of AGN selected mostly by means of their optical colors, and since we are only considering variability features in our selection, the confusion matrix and the different scores, obtained from our labeled sample, might not necessarily be an optimal prediction of the performance of our method in the unlabeled sample.

\begin{table}
\caption{Scores measured in the test set for each classifier} \label{tab:scores} 
\begin{center}

\begin{tabular}{ccccc} \hline

\hline

Score & RF1 & RF2 & RF3  \\
\hline

Accuracy & 0.916 & 0.923 & 0.931     \\
Precision & 0.909  & 0.909 &  0.921\\
Recall & 0.933 & 0.950 & 0.951 \\
F1 & 0.921 & 0.930 & 0.936  \\

 \hline
 \hline

\end{tabular}


\end{center}

\end{table}

\begin{figure}
\begin{center}
\includegraphics[scale=0.36]{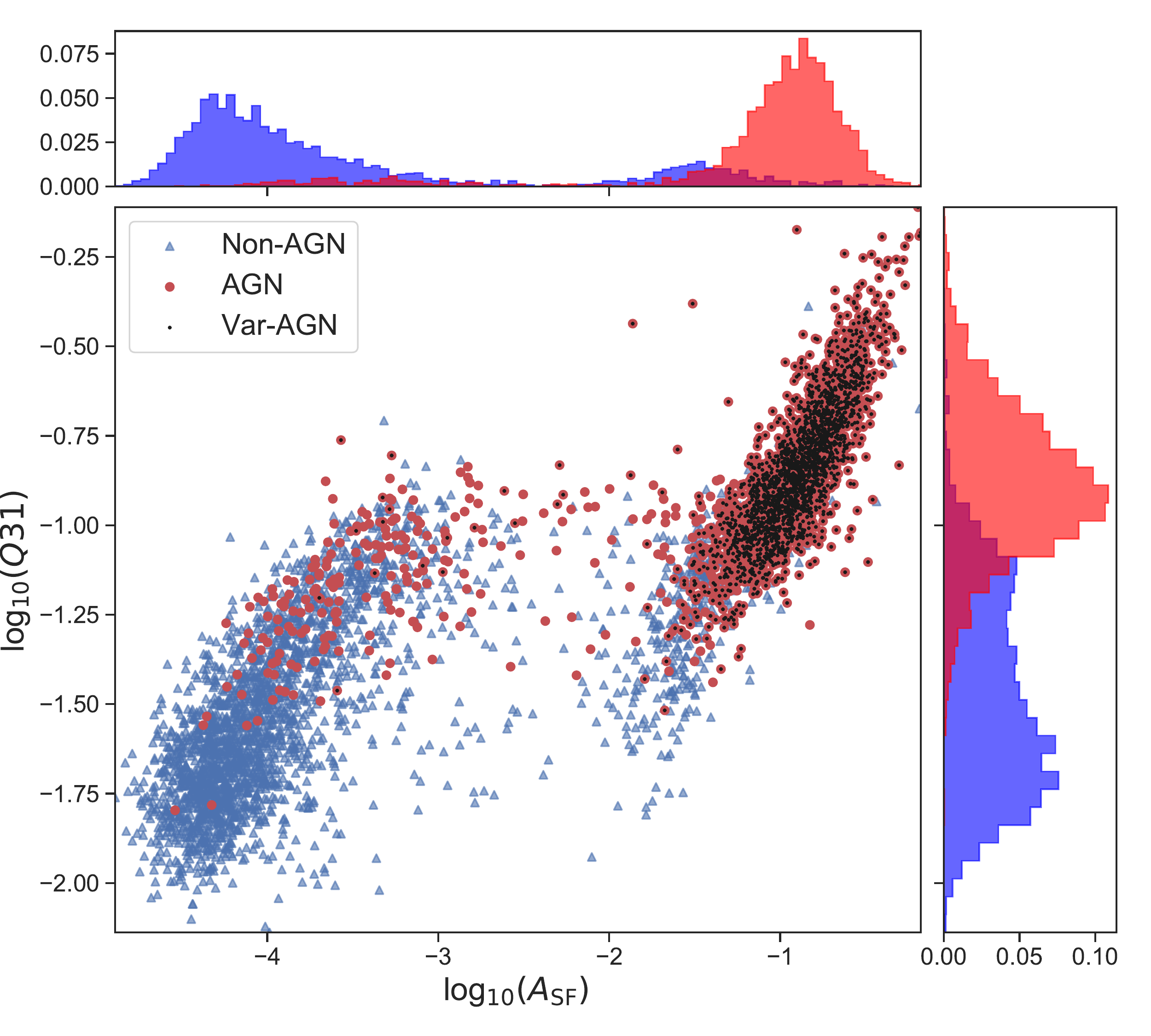}
\caption{Distribution of the  $A_\text{SF}$ and Q31 features for the labeled set. Blue triangles correspond to non--AGN, and red circles correspond to AGN. We mark with black dots those AGN classified as variable, according to the definition used in \cite{Sanchez17}. \label{figure:feat_training}}
\end{center}
\end{figure}

One of the advantages of the RF classification is that we can easily know the feature importance, since it provides a ranking score for each feature, or how well every feature separates the two classes. In the first columns of Table \ref{tab:features_ranking}, we provide the list of features, ordered by importance (rank value), for the RF1 classifier. It can be seen that the four most important features are the amplitude of the structure function, the excess variance, the Meanvariance, and Q31. In Figure \ref{figure:feat_training}, we show the distribution of the $A_\text{SF}$ and Q31 features for the labeled set. We highlight using black dots those AGN classified as variable, according to the definiton proposed by \cite{Sanchez17}, where a source is classified as variable when its light curve satisfies $P_{var} \geq 0.95$ and $(\sigma^2_{rms}-err(\sigma^2_{rms}))>0$. From the figure, it can be seen that AGN and non--AGN are separated by these two features, with $A_\text{SF}$ providing a much stronger division than Q31, as there is substantial source overlap between the two classes with the latter indicator. It can be also seen that the majority of the AGN with low variability amplitude are classified as non--variable.


\begin{table*}
\caption{Feature importance for each classifier.} \label{tab:features_ranking} 
\begin{center}

\begin{tabular}{lll|llll|llll} \hline
\hline

\multicolumn{2}{c}{RF1} & & & \multicolumn{2}{c}{RF2}  & & & \multicolumn{2}{c}{RF3}  \\

Feature &  Rank & & & Feature &  Rank  & & & Feature &  Rank    \\
\hline

$A_\text{SF}$  &   0.197    & & &  $A_\text{SF}$   & 0.209  & & & $A_\text{SF}$ &  0.189    \\
$\sigma_{\text{rms}}$   &   0.139 & & & $\sigma_{\text{rms}}$ & 0.149 & & & $\sigma_{\text{rms}}$ & 0.142 \\
Meanvariance  &   0.127 & & & Q31 & 0.102 & & & Q31 &  0.113  \\
Q31  &   0.111 & & & $P_{var}$ & 0.093 & & & Meanvariance &  0.099    \\
$P_{var}$  &   0.095 & & &  Std & 0.088  & & & $P_{var}$ & 0.095  \\
Std  &   0.090 & & & Meanvariance & 0.086 & & & Std &  0.074 \\
PercentAmp &   0.040 & & & PercentAmp  & 0.045 & & & Autocor--length & 0.042   \\
$\gamma_\text{SF}$  &   0.036  & & & Autocor--length & 0.035  & & & PercentAmp &  0.039 \\
Autocor--length &   0.033 & & &$\gamma_\text{SF}$ & 0.031   & & & $g-r$ & 0.031  \\
MedianBRP &   0.025 & & & $r-i$ & 0.028  & & & $\gamma_\text{SF}$  & 0.029  \\
LinearTrend   &   0.023 & & & MedianBRP & 0.021  & & & $r-i$ & 0.023  \\
PeriodLS   &    0.023 & & & PeriodLS & 0.020 & & & MedianBRP& 0.020  \\
$\eta^e$ &   0.023 & & & $\eta^e$ & 0.019  & & &$\eta^e$ & 0.020 \\
Beyond1Std &    0.019  & & & Beyond1Std  & 0.019 & & &  $i-z$ & 0.019 \\
StetsonK   &   0.018 & & & LinearTrend & 0.019 & & & LinearTrend  & 0.018 \\
Con &   0.002 & & & $i-z$ & 0.019 & & & PeriodLS & 0.017  \\
& & & & StetsonK   &  0.014 & & & Beyond1Std &   0.017 \\
& & & & Con & 0.002 & & &  StetsonK & 0.012 \\
& & & & & & &  & Con & 0.002 \\

\hline
 \hline

\end{tabular}


\end{center}

\end{table*}

\subsubsection{RF2: selection of AGN based on variability and $r-i$, and $i-z$ optical colors }

Our second RF classifier (RF2) includes variability features and the $r-i$ and $i-z$ colors. Figure \ref{figure:conf_matrix_all} shows the confusion matrix for RF2 (see its RF2 results). In this case, the confusion matrix is similar to the confusion matrix of RF1, however, in the case of RF2 we have a slightly cleaner population of AGN candidates (i.e. the fraction of False Positives is lower). The accuracy, precision, recall, and F1 scores are given in Table \ref{tab:scores}. There are not significant differences between the score values of RF1 and RF2.

In Table \ref{tab:features_ranking} we list the ranking of features for the RF2 classifier. There are not significant differences when compared to RF1, and notably we found that variability features are more relevant for AGN selection than the $r-i$ and $i-z$ colors. In this case, the most important features are $A_\text{SF}$, $\sigma_{\text{rms}}$, Q31, and $P_{var}$. We also found that the $r-i$ color seems to be more relevant than the $i-z$ to classify our sources. 

\subsubsection{RF3: selection of AGN based on variability and $g-r$, $r-i$, and $i-z$ optical colors}

Our third RF classifier (RF3) includes variability features and the $g-r$, $r-i$, and $i-z$ colors. Figure \ref{figure:conf_matrix_all} shows the confusion matrix for RF3 (see its RF3 results). In this case, the fraction of True Positives (True AGN classified as AGN) is slightly higher than those of RF1 and RF2. However, we must consider that most of the AGN in the labeled set have been selected by means of their optical colors, which might explain the improvement of the results over the test set compared with the RF2 classifier. 

The scores for RF3 are listed in Table \ref{tab:scores}. In comparison to RF1 and RF2, the scores are slightly higher, particularly the precision. In Table \ref{tab:features_ranking}  we list the ranking of features for the RF3 classifier. In this case the most important features are $A_\text{SF}$, $\sigma_{\text{rms}}$, Q31, and the Meanvariance. The most important color is $g-r$, which is expected due the distribution of non--AGN and AGN in Figure \ref{figure:train_color}. It is well--known that much of the discriminating power for selecting unresolved AGN is in $u-g$ (e.g., \citealt{Braccesi70}), which is in agreement with our finding that $r-i$ and $i-z$ colors are not as relevant for AGN selection.

\subsection{AGN candidates from QUEST--La Silla}\label{candidates}

We applied the trained RF1, RF2, and RF3 classifiers to our unlabeled well--sampled set of 208,583 light curves. In order to improve the purity of our selection, we considered the predicted class probability $P_{\text{RF}}$ (computed as the mean predicted class probabilities of the trees in the forest) to select the final set of AGN candidates. We defined two samples of AGN candidates: a) the full-AGN sample, consisting of all sources classified as AGN by the RF classifier ($P_{\text{RF}}\geq 0.5$), and b) the hp--AGN sample, consisting in sources that have a high probability ($P_{\text{RF}}\geq 0.8$) of being an AGN based on the RF classifier.  In Table \ref{table:cand_summary} we provide a summary with the number of sources classified as AGN in both samples, for each classifier. For the case of the RF1 classifier, there are 17,120 sources in the full--AGN sample, and 5,941 sources in the hp--AGN sample. For the RF2 classifier there are 15,100 sources in the full--AGN sample, and 5,252 sources in the hp--AGN sample. Finally, for RF3, there are 13,810 sources in the full--AGN sample, and 4,482 sources in the hp--AGN sample. There are 4,054 candidates in common between the RF1, RF2, and RF3 hp--AGN samples. For the rest of the analysis we will only consider the  hp--AGN samples of each classifier.

 \begin{table*}
\caption{Number of AGN candidates per field, for each classifier} \label{table:cand_summary} 
\begin{center}

\begin{tabular}{cc|cccc|cccc|ccc} \hline
\hline

& & & \multicolumn{2}{c}{RF1}  & & & \multicolumn{2}{c}{RF2} & & & \multicolumn{2}{c}{RF3}  \\

Field  & & &  full--AGN &  hp--AGN & & &  full--AGN &  hp--AGN & & &  full--AGN &  hp--AGN    \\

\hline

COSMOS & & &  3,968 & 1,503 & & &   3,562 & 1,201 & & & 3,424 & 1,018  \\
XMM--LSS & & &  6,441 & 2,374 & & &   5,774 & 2,106 & & & 5,516 & 1,879   \\
Elais--S1 & & &  3,374  & 988  & & &   2,936 &  942 & & & 2,441 & 777 \\
ECDF--S & & &  3,337 & 1,076 & & &    2,828 & 1,003 & & & 2,429 & 808   \\

\hline

Total & & &  17,120 & 5,941 & & &   15,100 & 5,252 & & & 13,810 & 4,482   \\

 \hline
 \hline

\end{tabular}


\end{center}

\end{table*}

\begin{figure}
\begin{center}
\includegraphics[scale=0.6]{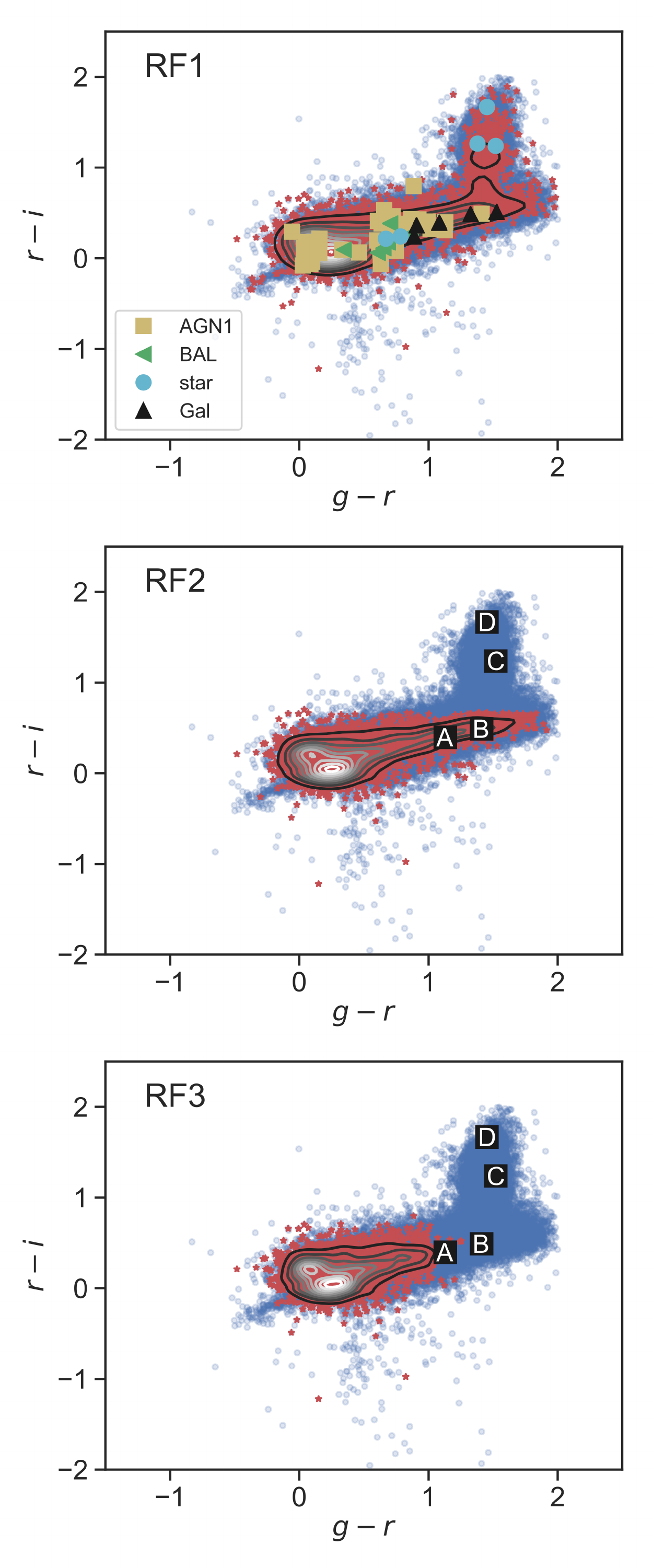}
\caption{$g-r$ vs $r-i$ color--color diagrams of the unlabeled set (blue circles), and the hp--AGN sample (red stars) for the RF1 (top panel), RF2 (middle panel), and RF3 (bottom panel). The contour plots show the distribution of the hp--AGN samples for each classifier. In the top panel, we show the position of candidates observed during spectroscopic follow--up, differentiating between type 1 AGN (AGN1, yellow squares), BAL-QSO (BAL, red triangles), stars (cyan circles), and galaxies (Gal, black triangles). In the middle and bottom panels we show with letters and black squares the position of a selection of observed candidates located in the stellar locus.\label{figure:color_unlabeled}}
\end{center}
\end{figure}

Figure \ref{figure:color_unlabeled} shows the $g-r$ vs $r-i$ color--color diagram of the unlabeled set, and the hp--AGN samples for the RF1, RF2, and RF3. Comparing with Figure \ref{figure:train_color}, we can see that several of our AGN candidates are located in regions of the color--color space where AGN are not normally found, particularly for the case of RF1. The main difference between the candidates of RF1 and the rest of the classifiers, is the exclusion of sources in the color--color region where we normally find cool stars. For example, there are 4,890 candidates in common between RF1 and  RF2, and 1,051 RF1 candidates that are not candidates for RF2. 54.1\% of the former ones have $r-i>0.7$, where we expect to find mostly cool stars.  The main differences between the candidates of RF2 and RF3, is the exclusion of the redder candidates in the case of RF3 ($g-r\gtrsim1.0$). There are 4,178 candidates in common between RF2 and RF3, and 1,074 candidates in RF2 that are not candidates for RF3, with 70.5\% of these having $g-r>1.0$. 

\begin{figure}
\begin{center}
\includegraphics[scale=0.5]{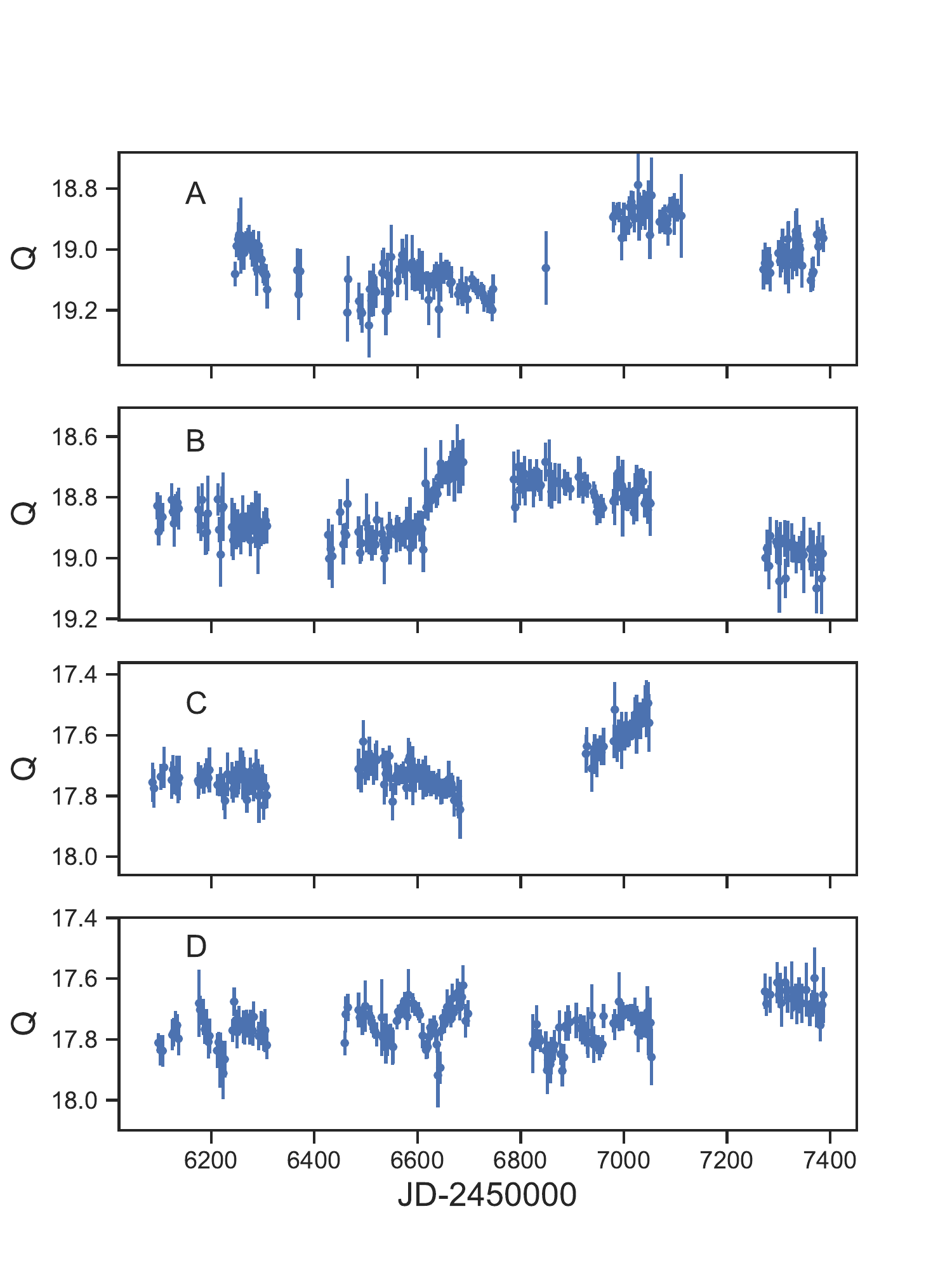}
\caption{Light curves of some RF1 candidates located in the stellar locus, observed during the spectroscopic follow--up campaign, shown in the top--right and bottom--left panels of Fig. \ref{figure:color_unlabeled}. A and B are classified as type 1 AGN, C and D are classified as M--type stars.\label{figure:lc_red}}
\end{center}
\end{figure}

In the top panel of Figure \ref{figure:color_unlabeled} we also show a selection of AGN candidates observed during spectroscopic follow--up (see Section \ref{follow_up}). In the middle and bottom panels of the figure we show the position in the $g-r$ vs $r-i$ diagram of four candidates located in different positions of the stellar locus, marked with letters (ABCD), and in Figure \ref{figure:lc_red} we show their light curves, where it can be seen that they are clearly variable. These four sources are selected as candidates by RF1, two are selected as candidates by RF2 (A and B), and none of them is selected as a candidate by RF3.

\section{Confirmation of AGN candidates}\label{cand_conf}

In the following sub--sections, we aim to confirm the nature of our candidates. In Section \ref{other_cat}, we use ancillary data to confirm the nature of our AGN candidates. In Section \ref{follow_up} we show the results of our spectroscopic follow--up campaign, conducted between December 2016 and September 2018, to test the efficiency of our selection method.

We are particularly  interested in identifying the nature of sources located in positions of the color--color space dominated by stars (e.g., in the stellar locus). We divided our high probability candidates according to their $g-r$ colors, we avoid $u-g$ since the $u$ band is not available for all the fields. We define the blue sub--sample as the one composed of sources with $g-r\leq0.6$ and the red sub--sample as that composed of sources with $g-r>0.6$. As can be seen in Figure \ref{figure:train_color}, most of the AGN in the labeled set have $g-r\leq0.6$.

\subsection{Confirmation by ancillary data}\label{other_cat}

MILLIQUAS  (v5.7 update, 7 January 2019, \citealt{Flesch15}) provides a very complete compendium of known AGN (both type 1 and type 2), from the literature, including the last data release of SDSS (SDSS--DR15), and several recent XMM--Newton, Swift, and Chandra catalogs (e.g., \citealt{Evans14,Marchesi16,Maitra19}). It also includes a list of  high--confidence  AGN candidates from different sources like AllWISE \citep{Secrest15}.

We used MILLIQUAS to confirm the nature of our candidates. We cross--matched MILLIQUAS with the coordinates of our well--sampled light curves, using a radius of 1\arcsec. There are 3,524 sources in the well--sampled sample with identifications in MILLIQUAS. For the case of the RF1 classifier, there are 2,358 (66.9\%) of these sources in the hp--AGN sample, 2,757 (78.2\%) in the full--AGN sample, and 767 (21.8\%) sources classified as non--AGN. For the RF2 classifier, there are 2,366 (67.1\%) sources in the hp--AGN sample, 2775 (78.7\%) in the full--AGN sample, and 749 (21.3\%) sources classified as non--AGN. Finally, for the RF3 classifier, there are 2,348 (66.7\%) sources in the hp--AGN sample, 2,769 sources in the full--AGN sample (78.6\%), and 755 (21.4\%) sources classified as non--AGN.  From these results we can say that $\sim 21\%$ of the sources are misclassified as non--AGN when we include variability features in the selection. Thus, we can conclude that when we include variability features in our selection, we obtain a completeness of $\sim 79\%$.

\begin{figure}
\begin{center}
\includegraphics[scale=0.6]{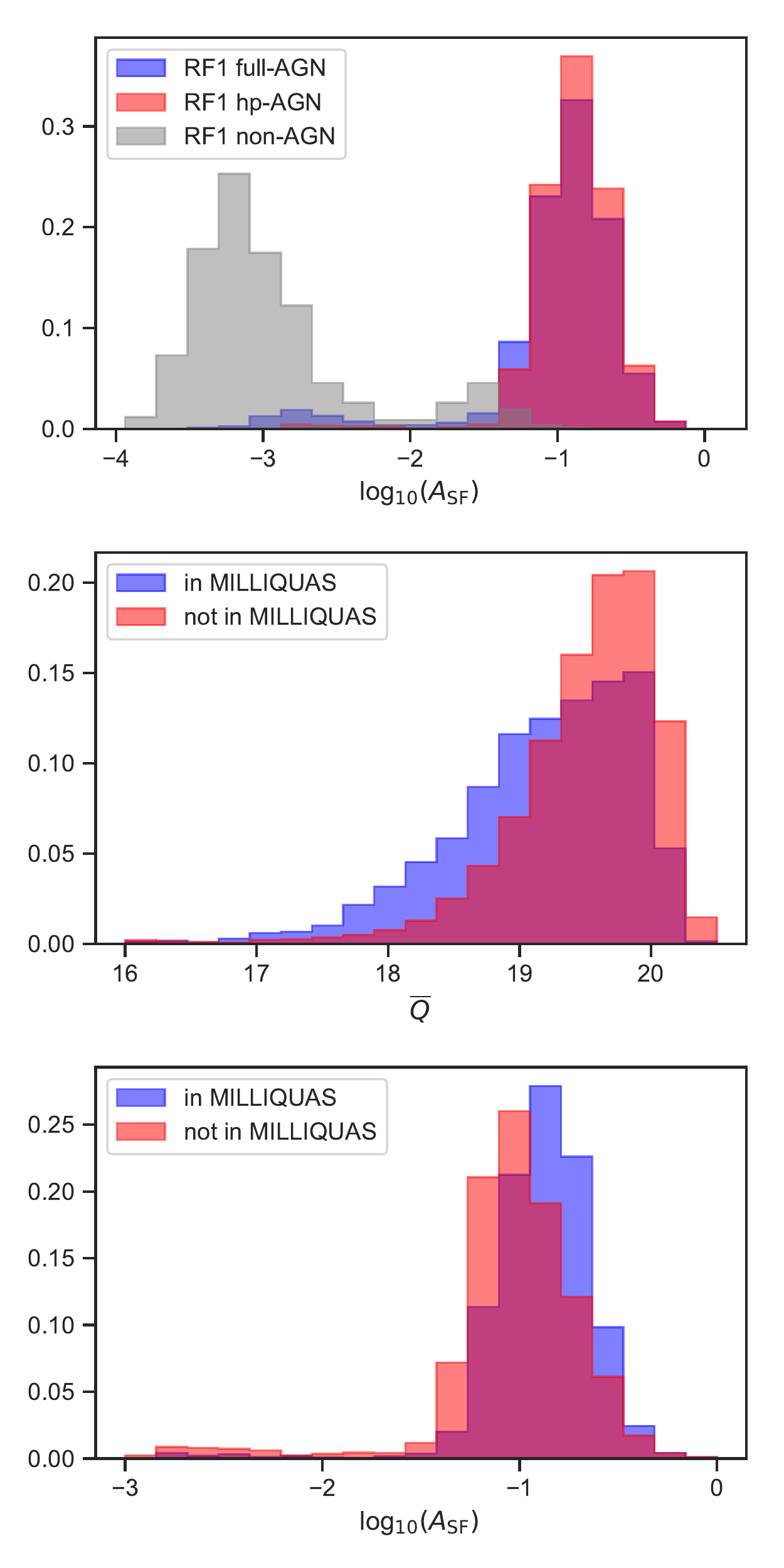}
\caption{Top panel: normalized distribution of $A_\text{SF}$ for sources with detection in MILLIQUAS with well--sampled light curves in QUEST--La Silla. We show sources from the RF1 full--AGN sample (blue), the RF1 hp--AGN sample (red), and sources classified as non--AGN by RF1. Middle panel: normalized distribution of the mean $Q$ magnitude for sources from the hp--AGN sample that are present (blue) and not present (red) in MILLIQUAS. Bottom panel: normalized distribution of $A_\text{SF}$ for sources from the hp--AGN sample that are present (blue) and not present (red) in MILLIQUAS. \label{figure:milliquas_agn}}
\end{center}
\end{figure}

We plot in the top panel of Figure \ref{figure:milliquas_agn} the distribution of the $A_\text{SF}$  variability feature, for sources in MILLIQUAS and QUEST--La Silla belonging to the RF1 hp--AGN sample, the RF1 full--ANG sample, and sources classified as non--AGN by RF1 (but classified as AGN by MILLIQUAS). It can be seen that the main difference between the sources classified as AGN and non--AGN is the value of the variability amplitude at one year, .i.e. we are not detecting variability for the sources classified as non--AGN. 

In the middle panel of Figure \ref{figure:milliquas_agn} we compare the distribution of the mean $Q$ magnitude (determined from their light curves) of sources from the RF1 hp--AGN sample that have and do not have detection in MILLIQUAS, and in the bottom panel we compare their $A_\text{SF}$ distributions. It can be seen that the sources with detection in MILLIQUAS are in general brighter than the sources without detection in MILLIQUAS, however the amplitude of the variability is lower for the sources without detection in MILLIQUAS. It is important to note that only a small fraction ($\lesssim 3\%$) of the AGN in MILLIQUAS are classified as host--dominated (i.e., they appear extended in imaging). This demonstrates that variability can be used to augment AGN selection to include objects that are extended as well as point sources. 

In Table \ref{table:cand_MILLIQUAS} we show the number of hp--AGN candidates confirmed using MILLIQUAS, dividing the hp--AGN samples into red and blue sub--samples, as already described. It can be seen that most of the confirmed sources are in the blue sub--sample, with 52\% of the candidates in this sample confirmed for RF1, 52.4\%  for RF2, and 52.1\% for RF3. For the case of the red sub--sample, 5.3\% of the candidates from RF1 are confirmed using MILLIQUAS, 7.1\% for RF2, and 12\% for RF3.  Besides, there are 354 AGN listed as candidates in MILLIQUAS for RF1, 356 for RF2, and 345 for RF3. Most of them are candidates from WISE \citep{Secrest15}. This lack of confirmed red candidates can be understood if we consider that most of the AGN presented in MILLIQUAS come from samples that applied morphological cuts to target point sources, thus, they tend to exclude sources whose emission is dominated by their host galaxies.

 \begin{table*}
\caption{Number of hp--AGN candidates confirmed using MILLIQUAS, for each classifier} \label{table:cand_MILLIQUAS} 
\begin{center}

\begin{tabular}{cc|cccc|cccc|ccc} \hline
\hline

& & & \multicolumn{2}{c}{RF1}  & & & \multicolumn{2}{c}{RF2} & & & \multicolumn{2}{c}{RF3}  \\

Sample  & & & blue & red & & & blue & red & & & blue & red \\
\hline
MILLIQUAS AGN & & &  1,882 & 122 & & & 1,894 & 116 & & & 1,901 & 100  \\
MILLIQUAS candidate & & & 324 & 30 & & & 322 & 34 & & & 320 & 25  \\
X--ray detections & & & 640 & 59 & & &  644 & 57  & & & 641 & 47 \\
hp--AGN & & & 3,618 & 2,323 & & & 3,613 & 1,639 & & & 3,646 & 836 \\

 \hline
 \hline

\end{tabular}


\end{center}

\end{table*}

\subsubsection{Candidates with X-ray detections}\label{xray_cat}

MILLIQUAS provides X--ray detections associated to every source, however some recent catalogs like \cite{Luo17} \cite{Chen18} are not completely included. Thus, we used different X--ray catalogs to complement the information provided by MILLIQUAS and see which candidates have X-ray detections associated. For the COSMOS field we used the optical and infrared counterparts catalog of the Chandra COSMOS-Legacy Survey \citep{Marchesi16}, for the XMM--LSS field we used the recent XMM-SERVS survey catalog \citep{Chen18}, for the ECDF--S field we used the Chandra Deep Field-South 7 Ms source catalog \citep{Luo17}, and for the Elais--S1 field we used Elais--S1 field X-ray source optical/IR Identifications catalog \citep{Feruglio08}.
 
In table \ref{table:cand_MILLIQUAS} we provide the number of candidates with X--ray detections from the previously mentioned X--ray catalogs or  MILLIQUAS (see the row ``X--ray detections''). It can be seen that most of the candidates with X--ray detections are from the blue sub--sample.  

\subsection{Spectroscopic follow--up of AGN candidates}\label{follow_up}

Since most of the AGN confirmed with ancillary data have blue colors ($g-r\leq0.6$), we performed spectroscopic follow--up to confirm the nature of sources located in different regions of the color--color space. We used Goodman at SOAR  \citep{SOARGoodman} and EFOSC2 at NTT \citep{NTTEFOSC2} instruments to observe 54 candidates (for details see Section \ref{spec_class} of the appendix). 

To select the candidates for the follow--up campaign we divided the hp--AGN sample of the RF1 classifier into the blue and red sub--samples. Then, we randomly selected 100 candidates from each sub--sample, excluding sources with $r>20.5$ for which it would be hard to obtain a good--quality spectrum with 4--meter class telescopes. We visually inspected the light curves of the selected candidates in order to exclude sources with evidence of bad photometry (produced by the relatively low cosmetic quality of the QUEST CCD camera chips). During the follow--up campaign we observed as much of the selected candidates as we could,  observing in total 54 targets. We gave priority to sources from the red sub--sample. Of the 54 candidates observed, 38 have $g-r > 0.6$, which represents 70\% of the sample.

\begin{figure}
\begin{center}
\includegraphics[scale=0.6]{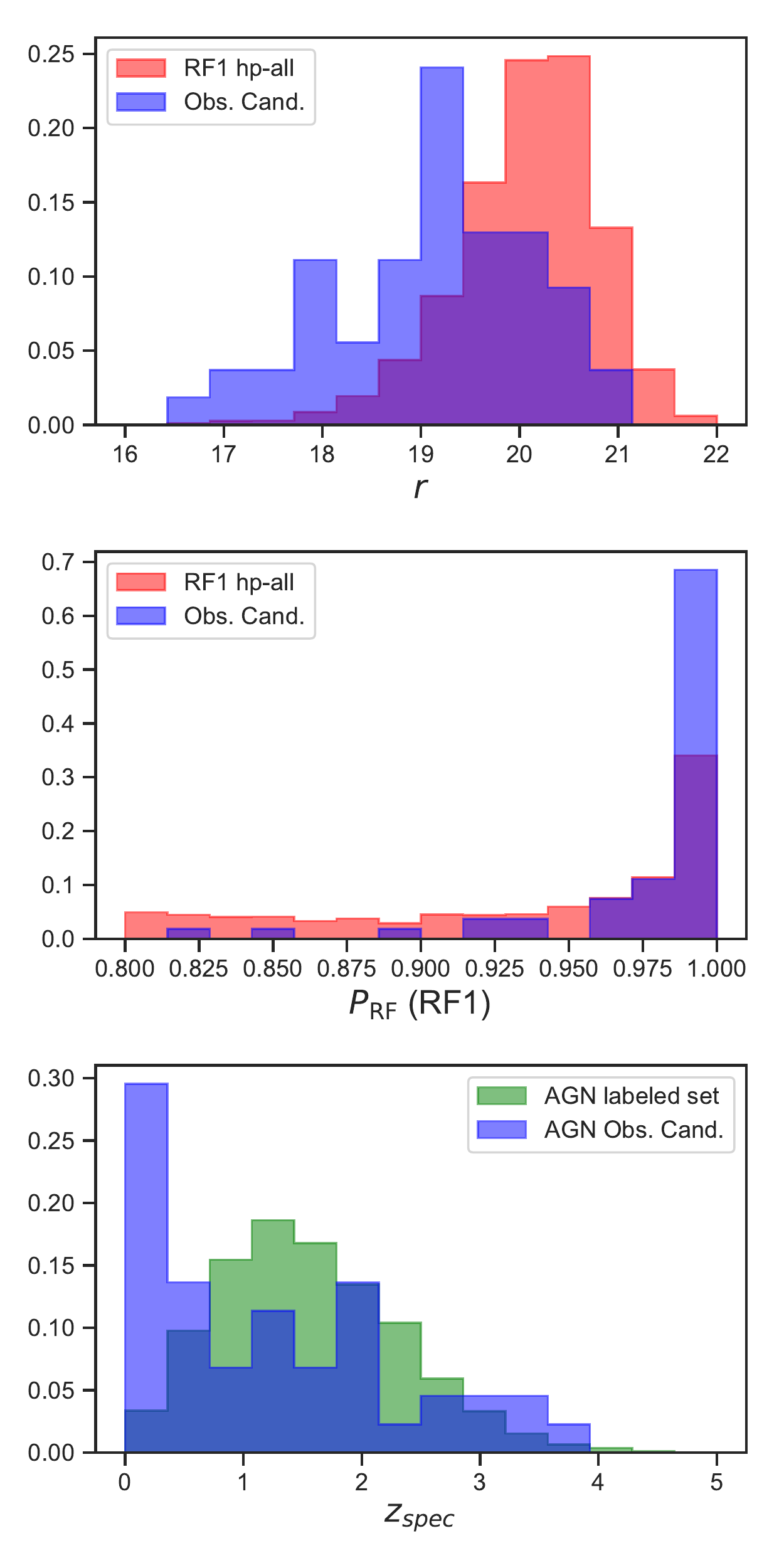}
\caption{Top panel: normalized histogram of the $r$ band magnitude of the RF1 hp--AGN sample (red) and the observed candidates (blue). Middle panel: normalized histogram of the predicted class probability ($P_{\text{RF}}$) of the RF1 hp--AGN sample (red) and the observed candidates (blue). Bottom panel: normalized histogram of spectroscopic redshift of AGN from the labeled sample (green) and observed candidates classified as type 1 AGN or BAL--QSO (red). \label{figure:rmag_proba}}
\end{center}
\end{figure}

In the top panel of Figure \ref{figure:rmag_proba} we show the $r$ band magnitude distribution of hp--AGN candidates of the RF1 classifier and the observed candidates. We can see that the distributions are different. This is produced by the limitations of observing faint targets with 4--meter class telescopes. During the follow--up campaign we gave priority to sources with $r<20$, for which we expected to obtain spectra with signal to noise higher than 10. In the bottom panel of  Figure \ref{figure:rmag_proba}  we show the distribution of the predicted class probability ($P_{\text{RF}}$) for the hp--AGN candidates of the RF1 classifier and the observed candidates. It can be seen that in general we observed sources with higher probabilities, compared with the total sample of candidates. This is produced by the visual inspection of the light curves and the selection of brighter sources for the follow--up campaign.

We used the spectra to classify our targets and to estimate their redshifts. For details about the spectroscopic analysis see section \ref{spec_class} of the appendix. The full list of observed candidates can be found in Section \ref{catalog} of the appendix.  We provide the position of the observed sources, their redshift, their $g-r$ and $r-i$ colors, their $r$ magnitude, and their spectroscopic classification. In Table \ref{table:summary_follow_up} we provide a summary of the follow--up campaign. We divided the classified sources into blue and red sub--samples, and we also separate them according to their spectroscopic classes: AGN1 (type 1 AGN), BAL--QSO, galaxy, and star. 

In the top--left panel of Figure \ref{figure:color_unlabeled}, we show the color--color diagram of the observed candidates, with colors and shapes depending on their spectral classification. In the top--right and bottom--left panels of Figure \ref{figure:color_unlabeled} we mark with letters some candidates located in the stellar locus. The sources A and B are classified as type 1 AGN, and the sources C and D are classified as type M stars. From the light curves of sources C and D (see Figure \ref{figure:lc_red}) and from their spectra, we propose that these candidates are irregular variable stars. 

In the bottom panel of Figure \ref{figure:rmag_proba} we show the normalized redshift distribution of AGN from the labeled sample, and observed candidates classified as AGN or BAL--QSO. We can see that the we have a much larger fraction of low--redshift sources observed during the follow--up campaign. Besides, the fraction of observed AGN with $z>3.0$ is slightly higher compared with the AGN from the labeled sample.

 \begin{table}
\caption{Summary of the spectroscopic follow--up campaign} \label{table:summary_follow_up} 
\begin{center}

\begin{tabular}{cc|cccc|cc} \hline
\hline

Class  & & &  blue sub--sample & red sub--sample & && Total  \\

\hline

AGN1 & & & 15 & 25 & && 40  \\
BAL--QSO& & & 1 & 3 & && 4  \\
Galaxy & & & 0 & 5 & && 5  \\
Star & & & 0 & 5& && 5  \\

 \hline
 \hline

\end{tabular}


\end{center}

\end{table}

There are seven targets with redshift higher than 2.5. These types of AGN are harder to detect than lower--redshift AGN in magnitude--limited, optical color--color selections (because their colors resemble those of stars, particularly near the magnitude limit of surveys where the stellar locus is wider), and clearly benefit from the variability criteria \citep{PalanqueDelabrouille11,Butler11,PalanqueDelabrouille16}. In addition, we found four BAL--QSO, with three having $g-r>0.6$. There are 22 AGN with $z_{spec}<0.7$, of which 20 have $g-r>0.6$, and eight of these are LLAGN, whose continua are significantly dominated by the host galaxy, but with clearly distinguishable broad emission lines.

\begin{figure}
\begin{center}
\includegraphics[scale=0.6]{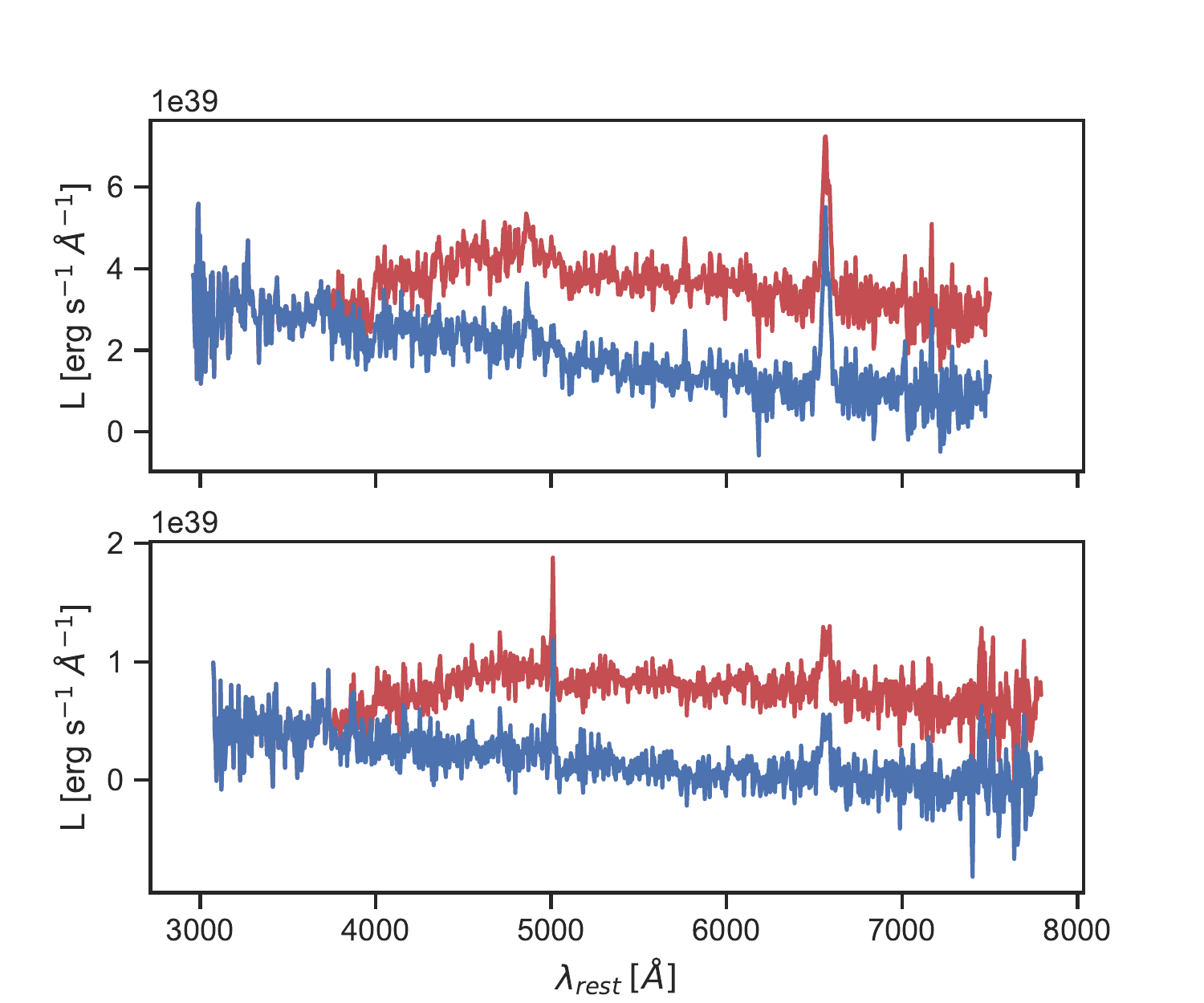}
\caption{Rest--frame optical spectra of two type 1 AGN with evidence of continuum dominated by the host galaxy (LLAGN), observed with EFOSC2/NTT. In red we show the original spectra, and in blue we show the AGN component. The most prominent emission lines correspond to H$\alpha$. \label{figure:red_spec}}
\end{center}
\end{figure}

The case of the eight LLAGN is particularly interesting, since the continuum of their spectra is dominated by the host galaxy, but we still detect its optical variable component, which is associated with the accretion disk. This component is revealed by subtracting the galactic continuum of each spectrum following the simple procedure of \cite{Greene05} and \cite{Kim06}. As an example, in Figure \ref{figure:red_spec} we show the spectra of two LLAGN sources, in red we show the original observed spectra and in blue we show the AGN component. It can be seen that in both cases the continuum is dominated by the host galaxy, but after its subtraction, the power--law AGN continuum appears, which is the one that produces the optical variations. It is important to remark that without including variability features in the selection of our candidates, these type of sources would be classified as non--AGN according to their optical colors. This reflects the importance of including variability in the selection of LLAGN.

We define the efficiency of a classifier as the number of confirmed AGN divided by the total number of observed candidates. Considering all the observed candidates, the efficiency is 100\% for the blue sub--sample, and 73.7\% for the red sub--sample. In Table \ref{table:follow_up} we show from which classifier comes every observed candidate. As we mentioned previously, the 54 observed candidates belong to the RF1 hp--AGN sample, therefore the efficiency of the follow--up for the RF1 classifier is 100\% for the blue sub--sample, and 73.7\% for the red sub--sample. For the case of RF2, there are 50 observed candidates, with an efficiency of 100\% for the blue sub--sample and 79.4\% for the red sub--sample. There are three stars and one type 1 AGN  excluded by RF2. For the case of RF3, there are 43 observed candidates. The efficiency of the RF3 blue sub--sample is 100\% and for the red sub--sample is  85.2\%. There are four type 1 AGN, one BAL--QSO, three galaxies, and three stars excluded by RF3. 

From these results, we can conclude that RF2 has a higher efficiency compared to RF1 and RF3, since it has a high efficiency for both the blue and red sub--samples, and excludes most of the observed stars and only one type 1 AGN. RF3 also provides good results, however it excluded one BAL--QSO and four type 1 AGN, of which one has $z_{spec}=3.5852$, and two are LLAGN. In Appendix \ref{RF2_cand} we provide the list of hp--AGN candidates from the RF2 classifier.

\subsubsection{Non--AGN observed sources}\label{contaminators}

We observed 10 sources which were spectroscopically classified as stars  or galaxies, but as AGN by our classifiers; all of which have clear evidence of variability. These 10 sources are selected as candidates by RF1, seven are selected as candidates by RF2 (two star and five galaxies), and four are selected as candidates by RF3 (two stars and two galaxies).

For the galaxy--classified cases, the light curves show clear signs of AGN--like variability. The obtained spectra of these sources are generally noisy (signal to noise less than 10), so we could be missing some weak emission lines. We tried to subtract a galactic component from these sources, and in some cases we found evidence of a power--law continuum component, but without evidence of emission lines. We decided to classify these sources as galaxies, however, in order to confirm their true nature, we likely need deeper observations, using 8--meter class telescopes.  We note that \cite{Cartier15} found that about 20\% of the objects classified spectroscopically as galaxies showed variability, they found a similar percentage of  narrow--line AGN showing variability.

For the star--classified cases, four of them are M--type stars, and one seems to be a K--type star. From their light curves, we can conclude that they are semi--periodic or irregular variable stars.

\section{Comparison with previous works}\label{comparison}

\cite{Butler11} used SDSS photometry to select AGN candidates through variability in the Stripe 82 field. They used damp random walk modelling \citep{Kelly09} to detect quasar--like variable sources. The light curves used in their analysis have on average 10 epochs, with a maximum of 28, obtained over $\sim 6$ years \citep{Sesar07}. As can be seen in Figure 8 of \cite{Butler11}, most of their candidates lie in the region where typical AGN are found, and only $\sim 1\%$ of their candidates lie in the color--color space dominated by stars (stellar locus). 

\cite{PalanqueDelabrouille11} and \cite{PalanqueDelabrouille16} used SDSS photometry to select AGN candidates through variability in the Stripe 82 field, to be observed as part of the BOSS and eBOSS surveys. On average, the light curves used by them had $53 \pm 20$ epochs, and a total coverage spanning 4 and 10 years. They characterized the variability of each source using the structure function, and they classified the sources using a neural network algorithm. They demonstrated that their method is very efficient at selecting sources with $z_{spec}>2.2$ or a BAL--QSO classification. However, the fraction of candidates with stellar--like colors is low (as can be seen in Figure 18 of \citealt{PalanqueDelabrouille11}).

\cite{Peters15} used SDSS data to perform a NBC KDE algorithm that classifies type 1 quasar, using color, variability and astrometric parameters. They used data in five broad optical bands ($u$, $g$, $r$, $i$, and $z$), and constructed light curves in each band with 10 to $\sim100$ observations over time--scales from $\sim$1 day to $\sim$8 years. They used structure functions to characterize the variability of each source. They tested different combinations of these parameters, finding that by combining variability and colors, they can achieve 97\% efficiency, improving particularly the efficiency in the selection of quasars at $2.7 < z < 3.5$. They selected 35,820 type 1 quasar candidates, with only the 14\% of them having $g-r>0.6$.

More recentrly, \cite{Tie17} combined colors and variability properties to select AGN candidates from DES. They obtained light curves from DES, which span less than a year and typically have $\sim 15$ epochs.  They used the chi--squared integrated probability to select AGN candidates. Since they have light curves with only a year of coverage, they did not implement more sophisticated variability selection methods. They demonstrated that combining variability with optical and infrared photometry improves the efficiency of AGN selection. \cite{Tie17} provide a catalog of 1,263 spectroscopically confirmed quasars in the DES supernova fields brighter than i = 22 mag. Only 6\% of their confirmed candidates have $g-r>0.6$. 

The light curves used in our analysis have a considerably higher cadence than the ones used in previous variability analyses, with an average of $119\pm47$ epochs and a total length of $1306\pm254$ days. In our case, we find that 39.1\%, 31.2\%, and 18.7\% of the hp--AGN candidates from RF1, RF2, and RF3, respectively, have $g-r>0.6$, where stars outnumber AGN. As we show in Section \ref{follow_up}, most of the atypical AGN observed during our spectroscopic follow--up campaign lies is this region of color--color space (70\% of the observed candidates). In general, our selection technique does not differ considerably to other techniques (e.g., \citealt{PalanqueDelabrouille11,Peters15}). The key difference is the larger number of epochs and the larger coverage of the QUEST--La Silla light curves, in addition to the exclusion of morphological parameters during the selection of candidates. This help us to be more sensitive to atypical AGN, like BAL--QSO and LLAGN (observing four and eight, respectively, during the follow--up) compared with previous analyses, which might be related with the higher probability to detect a variable signal from our better quality light curves.

Our selection technique has the advantage of being easily applicable to LSST data, since the expected cadence for the DDF will be similar to the one used here \citep{LSSTbook}. However, it is important to note that LSST will provide $ugrizy$ photometry, with single--epoch depths reaching to $\sim$24th magnitude, and $\sim$27th magnitude for the stacked images. Therefore, LSST will allow to perform variability analyses of much fainter sources than the QUEST--La Silla survey, and with multi--band light curves, which can help to weed out false positives and allow for a more complete characterization of the variability properties of the candidates.

Moreover, LSST will have the advantage of including the $u$ band, which has been repeatedly used in the past to select AGN candidates (in combination with other photometric bands), after a morphological cut is applied (e.g., \citealt{Fan99,Richards02,Richards04,Smith05,Richards09,Kirkpatrick11,Bovy11,Ross12}). These color--color selection techniques will be easier to apply to the full stacked depths of the LSST data, where the variability--based methods (like those presented in here) will eventually become infeasible, due to large flux errors on the light curves. However, for the case of AGN with strong host contamination  variability--based methods will be in advantage over color--color selection techniques.

\section{Conclusions}\label{discussion}

We have presented a methodology to classify AGN through variability analyses, particularly useful to find AGN populations missed by other optical selection techniques. We used data from the QUEST--La Silla AGN variability survey to construct a total of 208,583 well--sampled light curves in the COSMOS, XMM--LSS, Elais--S1, and ECDF--S fields. We characterize the variability of these sources by using different variability features (see Section \ref{var_feat}). We used a Random Forest algorithm to classify our objects as either AGN or non--AGN using variability features and optical colors. We tested three classification schemes, one that includes only variability features (RF1), one that includes variability features and the $r-i$ and $i-z$ colors (RF2), and one that includes variability features and the $g-r$, $r-i$, and $i-z$ colors (RF3). We have a total of 5,941 AGN candidates for the RF1 classifier, 5,252 candidates for the RF2 classifier, and 4,482 candidates for the RF3 classifier.

We confirmed the nature of our candidates by using ancillary data, and we found that a high fraction of the candidates from each classifier with $g-r \leq 0.6$ are known AGN from the literature (52\% for RF1, 52.4\% for RF2, and 52.1\% for RF3; see Section \ref{other_cat}), but the fraction of candidates  with $g-r>0$ confirmed by ancillary data is low (5.3\% for RF1, 7.1\% for RF2, and 12\% for RF3). This is produced because most of the AGN known from the literature are biased against bluer optical colors by their selection criteria. This motivated us to perform spectroscopic follow--up, to confirm the nature of sources located in different regions of the color--color space.

We observed  54 candidates with EFOSC2/NTT and Goodman/SOAR, with 70\% of the observed targets having $g-r>0$. We confirm the nature of several interesting sources, including four BAL--QSOs, seven sources with $z_{spec}>2.5$, and eight LLAGN. Our method was very efficient in classifying AGN with $g-r \leq 0.6$, for which we achieved 100\% of efficiency for all the classifiers. For the case of sources with $g-r>0$, our method also demonstrated good performance, achieving 73.7\% efficiency for RF1, 79.4\% for RF2,  and 85.2\% for RF3.

From the spectroscopic follow--up campaign, we conclude that the optimal classifier is the one that includes variability features and the $r-i$ and $i-z$ colors (RF2), as it avoids the region of the color--color space where we normally find cool stars, and also shows high efficiency, excluding only one observed type 1 AGN. The RF3 classifier also provides good results, however it excluded four AGN and one BAL-QSO. For the case of RF1, we propose that most of the candidates with $g-r\sim 1.5$ and $r-i \gtrsim 0.8 $ are LPV or binary stars. 

Our work can be considered as a pilot study in preparation for LSST, since the selection techniques tested here can be easily implemented for LSST data. The cadence of the LSST's DDF will be similar to the one of QUEST--La Silla, but covering 10 years of observations, which will improve considerably the selection efficiency. In addition, LSST will provide observations in more than one photometric band, which should prove useful to discard artifacts and false positives. Particularly, LSST will provide $u$ band photometry, which has been exhaustively used in the literature for the selection of point--like AGN, with very high efficiencies (e.g., \citealt{Richards04,Richards09}). Thus, color--color selection methods  will remain a critical approach in the LSST era, particularly for the selection of faint, point--like AGN, since they can be applied to the full depths of the LSST data. However, optical color--color selections alone are not efficient at classifying morphologically extended AGN. For these type of objects, a combination of optical colors and variability--based methods will be more suitable, as we have demonstrated in this work.

\acknowledgments

We thank Patrick Hall for his help classifying some of the candidates observed during the spectroscopic follow--up. We also thank the referee for a careful reading of the manuscript and comments that led to its improvement. PS was supported by CONICYT through “Beca Doctorado Nacional, A\~no 2013” grant \#21130441. PS received partial support from Center of Excellence in Astrophysics and Associated Technologies (PFB 06). PL acknowledges Fondecyt Grant \#1161184. NM acknowledges the support of the Helmholtz Einstein International Berlin Research School in Data Science (HEIBRiDS), Deutsches Elektronensynchrotron (DESY), and Humboldt-Universit\"at zu Berlin.  LCH was supported by the National Key R\&D Program of China (2016YFA0400702) and the National Science Foundation of China (11721303). FEB acknowledges support from CONICYT-Chile (Basal AFB-170002) and the Ministry of Economy, Development, and Tourism's Millennium Science Initiative through grant IC120009, awarded to The Millennium Institute of Astrophysics, MAS. This work was partially funded by the CONICYT PIA ACT172033.

 This work was based on data products from observations made with ESO Telescopes at the La Silla Paranal Observatory under ESO program ID 0101.A-0417. This work is also based on observations obtained at the Southern Astrophysical Research (SOAR) telescope, which is a joint project of the Minist\'{e}rio da Ci\^{e}ncia, Tecnologia, Inova\c{c}\~{o}es e Comunica\c{c}\~{o}es (MCTIC) do Brasil, the U.S. National Optical Astronomy Observatory (NOAO), the University of North Carolina at Chapel Hill (UNC), and Michigan State University (MSU).

Funding for the SDSS and SDSS-II has been provided by the Alfred P. Sloan Foundation, the Participating Institutions, the National Science Foundation, the U.S. Department of Energy, the National Aeronautics and Space Administration, the Japanese Monbukagakusho, the Max Planck Society, and the Higher Education Funding Council for England. The SDSS Web Site is http://www.sdss.org/.

The SDSS is managed by the Astrophysical Research Consortium for the Participating Institutions. The Participating Institutions are the American Museum of Natural History, Astrophysical Institute Potsdam, University of Basel, University of Cambridge, Case Western Reserve University, University of Chicago, Drexel University, Fermilab, the Institute for Advanced Study, the Japan Participation Group, Johns Hopkins University, the Joint Institute for Nuclear Astrophysics, the Kavli Institute for Particle Astrophysics and Cosmology, the Korean Scientist Group, the Chinese Academy of Sciences (LAMOST), Los Alamos National Laboratory, the Max-Planck-Institute for Astronomy (MPIA), the Max-Planck-Institute for Astrophysics (MPA), New Mexico State University, Ohio State University, University of Pittsburgh, University of Portsmouth, Princeton University, the United States Naval Observatory, and the University of Washington.

\bibliography{bibliography}

\appendix

\section{Catalog of observed candidates}	\label{catalog}

Here we present the list of candidates observed during our spectroscopic follow--up campaign. We provide the equatorial coordinates in degrees (J2000), telescope used, classifier from which the candidate was selected, measured redshift, quality flag of the measured redshift (1: low--quality $z_{spec}$, 2: good--quality $z_{spec}$), the magnitude in the $r$ band, the $g-r$ color, the $r-i$ color, and the spectroscopic classification. For details about the spectral analysis of these targets see Section \ref{spec_class} of the appendix.



\begin{center}
\begin{longtable}{ccccccccccc}
\caption{Targets observed during spectroscopic follow--up. \label{table:follow_up}}\\
\hline
Name & RA & DEC & Telescope & Classifier & $z_{spec}$ & FLAG$_z$ & $r$ & $g-r$ &  $r-i$ & Class \\
\hline
\endfirsthead
\multicolumn{4}{c}%
{\tablename\ \thetable\ -- \textit{Continued from previous page}} \\
\hline
Name & RA & DEC & Telescope & Classifier & $z_{spec}$ & FLAG$_z$ & $r$ &$g-r$  & $r-i$& Class \\
\hline
\endhead
\hline \multicolumn{4}{r}{\textit{Continued on next page}} \\
\endfoot
\hline
\endlastfoot

QLS\_1 & 7.021016 & -45.806145 & SOAR & RF1/RF2 &  3.5852 & 1 & 20.14 & 1.13 & 0.31 & AGN1 \\ 
QLS\_2 & 7.263506 & -45.629417 & NTT & RF1/RF2/RF3 &  1.3796 & 2 & 20.72 & 0.63 & 0.09 & AGN1 \\ 
QLS\_3 & 7.323368 & -43.633305 & SOAR& RF1/RF2/RF3  &  0.3242 & 1 & 18.57 & 0.03 & -0.08 & AGN1 \\ 
QLS\_4 & 7.377026 & -46.529945 & NTT & RF1/RF2 &  0.2824 & 1 & 19.32 & 1.53 & 0.51 & Gal \\ 
QLS\_5 & 7.387083 & -43.664276 & NTT & RF1/RF2 &  0.2000 & 1 & 19.16 & 1.08 & 0.39 & Gal \\ 
QLS\_6 & 7.418616 & -42.320072 & NTT & RF1 &  0.0 & 2 & 17.88 & 1.52 & 1.24 & STAR \\ 
QLS\_7 & 7.419530 & -43.789948 & NTT & RF1/RF2 &  0.3912 & 2 & 18.91 & 1.12 & 0.34 & AGN1 \\ 
QLS\_8 & 7.820146 & -45.645706 & NTT & RF1/RF2 &  0.3123 & 2 & 19.07 & 1.41 & 0.49 & AGN1 \\ 
QLS\_9 & 7.970508 & -42.275764 & NTT & RF1/RF2/RF3  &  0.1847 & 2 & 18.02 & 0.99 & 0.41 & AGN1 \\ 
QLS\_10 & 8.304768 & -45.681595 & NTT & RF1/RF2/RF3  &  0.2676 & 2 & 18.73 & 0.92 & 0.43 & AGN1 \\ 
QLS\_11 & 8.348364 & -46.856922 & NTT & RF1/RF2/RF3  &  3.5297 & 2 & 19.90 & 0.75 & 0.08 & AGN1 \\ 
QLS\_12 & 8.629325 & -42.310108 & SOAR & RF1 &  0.0 & 2 & 20.36 & 1.38 & 1.26 & STAR \\ 
QLS\_13 & 8.787973 & -45.348194 & NTT & RF1/RF2/RF3  &  0.1469 & 2 & 17.53 & 0.64 & 0.38 & AGN1 \\ 
QLS\_14 & 9.407302 & -43.000004 & NTT & RF1/RF2/RF3  &  1.986 & 1 & 19.69 & 0.69 & 0.17 & AGN1 \\ 
QLS\_15 & 9.414984 & -43.422619 & SOAR & RF1/RF2/RF3  &  1.8265 & 2 & 20.45 & -0.06 & 0.29 & AGN1 \\ 
QLS\_16 & 10.021671 & -43.859173 & NTT & RF1/RF2/RF3  &  0.3710 & 2 & 19.28 & 0.86 & 0.32 & AGN1 \\ 
QLS\_17 & 10.097470 & -44.866116 & NTT & RF1/RF2/RF3  &  2.9609 & 1 & 19.94 & 0.70 & 0.38 & BAL-QSO \\ 
QLS\_18 & 10.225745 & -43.855934 & NTT & RF1/RF2/RF3  &  1.2671 & 1 & 20.49 & 0.63 & -0.07 & AGN1 \\ 
QLS\_19 & 10.758265 & -42.452019 & NTT & RF1/RF2/RF3  &  0.1000 & 1 & 19.90 & 0.89 & 0.24 & Gal \\ 
QLS\_20 & 10.866718 & -43.825359 & NTT & RF1/RF2/RF3  &  3.123 & 1 & 20.58 & 0.65 & 0.11 & AGN1 \\ 
QLS\_21 & 11.090293 & -43.665966 & NTT & RF1/RF2/RF3  &  0.0 & 2 & 17.12 & 0.78 & 0.24 & STAR \\ 
QLS\_22 & 30.591522 & -2.020991 & SOAR & RF1/RF2/RF3  &  2.0502 & 2 & 18.66 & 0.04 & 0.11 & AGN1 \\ 
QLS\_23 & 30.603312 & -1.752825 & NTT & RF1/RF2/RF3  &  0.2101 & 2 & 18.12 & 0.71 & 0.33 & AGN1 \\ 
QLS\_24 & 31.057844 & -2.953287 & SOAR& RF1/RF2/RF3  &  0.1500 & 1 & 19.26 & 0.91 & 0.36 & Gal \\ 
QLS\_25 & 31.167528 & -3.630512 & NTT & RF1/RF2 &  1.22 & 1 & 19.14 & 0.61 & 0.06 & BAL-QSO \\ 
QLS\_26 & 31.533081 & -2.510754 & SOAR & RF1/RF2/RF3  &  1.4319 & 1 & 18.12 & 0.09 & 0.06 & AGN1 \\ 
QLS\_27 & 32.158398 & -3.652800 & NTT & RF1 &  0.0 & 2 & 18.02 & 1.45 & 1.67 & STAR \\ 
QLS\_28 & 32.456287 & -3.651828 & SOAR & RF1/RF2/RF3  &  1.4976 & 2 & 18.85 & 0.09 & 0.14 & AGN1 \\ 
QLS\_29 & 33.409081 & -3.250347 & SOAR  & RF1/RF2/RF3  &  2.8491 & 2 & 19.54 & 0.16 & 0.06 & AGN1 \\ 
QLS\_30 & 33.474072 & -3.279924 & NTT & RF1/RF3  &  0.5671 & 1 & 20.81 & 0.88 & 0.80 & AGN1 \\ 
QLS\_31 & 36.426113 & -2.971209 & NTT & RF1/RF2/RF3  &  0.0 & 2 & 16.53 & 0.67 & 0.22 & STAR \\ 
QLS\_32 & 36.852539 & -2.401858 & NTT & RF1/RF2/RF3  &  0.2551 & 2 & 19.94 & 0.84 & 0.38 & AGN1 \\ 
QLS\_33 & 37.988422 & -2.585521 & SOAR  & RF1/RF2/RF3  &  0.3498 & 2 & 18.83 & 0.09 & 0.01 & AGN1 \\ 
QLS\_34 & 38.680885 & -2.637419 & SOAR  & RF1/RF2/RF3  &  0.8437 & 2 & 19.86 & 0.46 & 0.06 & AGN1 \\ 
QLS\_35 & 51.529488 & -29.656691 & NTT & RF1/RF2/RF3  &  0.2307 & 2 & 18.15 & 0.87 & 0.43 & AGN1 \\ 
QLS\_36 & 51.765114 & -27.740358 & SOAR  & RF1/RF2/RF3  &  2.0279 & 2 & 19.20 & 0.07 & 0.17 & AGN1 \\ 
QLS\_37 & 52.009411 & -28.600405 & NTT & RF1/RF2 &  0.2667 & 2 & 18.01 & 1.33 & 0.49 & Gal \\ 
QLS\_38 & 52.013538 & -30.619936 & NTT & RF1/RF2/RF3  &  0.3284 & 2 & 19.47 & 0.90 & 0.37 & AGN1 \\ 
QLS\_39 & 52.397503 & -27.657492 & NTT & RF1/RF2/RF3  &  1.4175 & 2 & 19.59 & 0.60 & 0.20 & AGN1 \\ 
QLS\_40 & 52.536362 & -29.822481 & NTT & RF1/RF2/RF3  &  0.1804 & 2 & 17.51 & 0.66 & 0.53 & AGN1 \\ 
QLS\_41 & 52.593563 & -29.353525 & NTT & RF1/RF2/RF3  &  0.2177 & 2 & 19.94 & 0.61 & 0.41 & AGN1 \\ 
QLS\_42 & 53.317341 & -29.488207 & SOAR  & RF1/RF2/RF3  &  1.9234 & 2 & 19.48 & 0.16 & 0.22 & AGN1 \\ 
QLS\_43 & 53.730831 & -27.736212 & NTT & RF1/RF2/RF3  &  3.4986 & 1 & 19.37 & 0.86 & 0.27 & BAL-QSO \\ 
QLS\_44 & 53.749317 & -27.499168 & NTT & RF1/RF2 &  0.3598 & 2 & 19.42 & 1.13 & 0.40 & AGN1 \\ 
QLS\_45 & 53.847992 & -28.123224 & SOAR  & RF1/RF2/RF3  &  0.8682 & 2 & 17.11 & 0.09 & -0.06 & AGN1 \\ 
QLS\_46 & 53.864979 & -26.950115 & SOAR  & RF1/RF2/RF3  &  0.8159 & 2 & 19.50 & 0.04 & 0.08 & AGN1 \\ 
QLS\_47 & 53.911106 & -28.961195 & NTT & RF1/RF2/RF3  &  0.2116 & 2 & 18.54 & 0.72 & 0.46 & AGN1 \\ 
QLS\_48 & 53.943211 & -27.432163 & NTT & RF1/RF2/RF3  &  0.4332 & 1 & 19.18 & 1.02 & 0.32 & AGN1 \\ 
QLS\_49 & 54.049484 & -28.095388 & SOAR  & RF1/RF2/RF3  &  2.5265 & 2 & 18.87 & 0.10 & -0.04 & AGN1 \\ 
QLS\_50 & 54.782047 & -26.617104 & SOAR  & RF1/RF2/RF3  &  0.3825 & 2 & 19.14 & 0.70 & 0.26 & AGN1 \\ 
QLS\_51 & 54.785744 & -28.131121 & SOAR  & RF1/RF2/RF3  &  2.1388 & 1 & 19.20 & 0.34 & 0.09 & BAL-QSO \\ 
QLS\_52 & 54.878654 & -27.307127 & SOAR  & RF1/RF2/RF3  &  1.0899 & 2 & 19.72 & 0.62 & 0.16 & AGN1 \\ 
QLS\_53 & 55.059608 & -26.485237 & SOAR  & RF1/RF2/RF3  &  1.6453 & 2 & 19.43 & 0.05 & 0.17 & AGN1 \\ 
QLS\_54 & 149.004532 & 1.161204 & SOAR  & RF1/RF2/RF3  &  2.3530 & 1 & 20.36 & 0.33 & 0.09 & AGN1 \\ 

\end{longtable}
\end{center}



\section{Spectroscopic analysis of the observed candidates}	\label{spec_class}

We obtained classification spectra for 21 of our candidates using both the red and blue cameras of the Goodman spectrograph \citep{SOARGoodman}, mounted on the SOAR telescope. We used the 400 lines mm$^{-1}$ grating and the 1.0\arcsec and 0.8\arcsec slits proving a typical resolution of  $\sim 6 \AA$ or better. We reduced Goodman data following usual steps including bias subtraction, flat fielding, cosmic ray rejection (see \citealt{vanDokkum01}), wavelength calibration, flux calibration, and telluric correction using our own custom IRAF\footnote{IRAF is distributed by the National Optical Astronomy Observatories, which are operated by the Association of Universities for Research in Astronomy, Inc., under cooperative agreement with the National Science Foundation.} routines.

We also obtained classification spectra for 33 candidates using EFOSC2 \citep{NTTEFOSC2}  mounted on the New Technology Telescope (NTT) at La Silla Observatory.  We used the 236 lines mm$^{-1}$ grating and the 1.0\arcsec slit providing a typical resolution of 18 \AA. We followed the same observing procedures as the Public ESO Spectroscopic Survey for Transient Objects (PESSTO) collaboration \citep{Smartt15}, using the PESSTO Observing Blocks (OBs) to perform our observations, We reduced our observations using the PESSTO pipeline \citep{Smartt15}.

\begin{figure*}
\begin{center}
\includegraphics[scale=0.5]{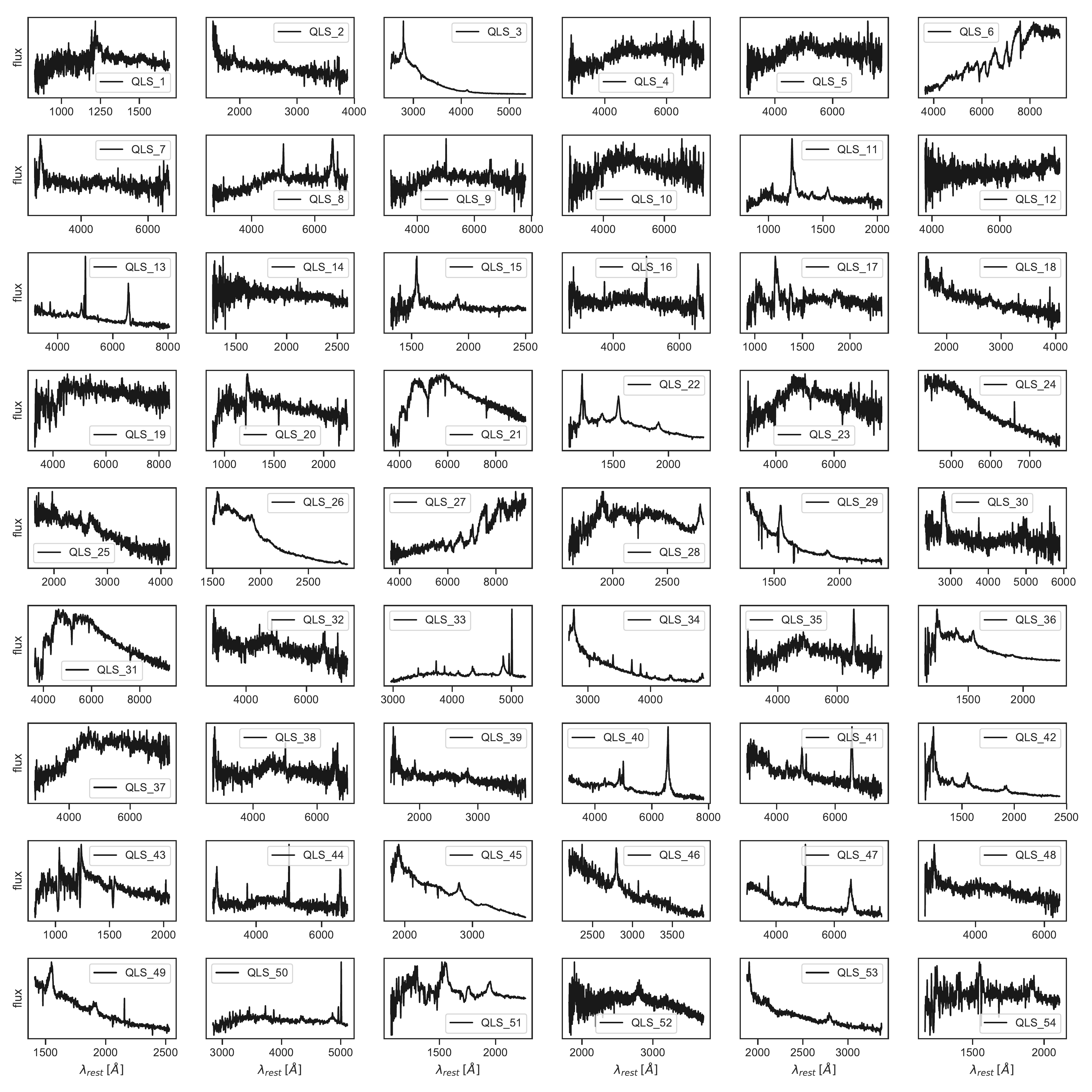}
\caption{Rest--frame optical spectra of the observed candidates. The flux is in arbitrary units. \label{figure:spec_all_A}}.
\end{center}
\end{figure*}

We corrected the reduced and calibrated one-dimensional spectra by Galactic extinction using the maps of \cite{Schlegel98} and the model of \cite{Cardelli89}. We then computed their redshifts and spectral classes by cross--correlating every spectrum with a set of spectral classification templates from SDSS\footnote{http://classic.sdss.org/dr5/algorithms/spectemplates/}, a type 1 AGN composite spectrum \citep{Croom02}, and a type 2 AGN spectrum \citep{Jones09}. We define a redshift FLAG that indicates the quality of the measured redshift. We say that a computed redshift has good--quality (FLAG=2) when there are several lines in the spectrum, and these lines are not affected by absorption features; and we say that a computed redshift has low--quality (FLAG=1) when the spectrum has low signal to noise, when the number of emission lines available is low (one or two), or when the emission lines are highly affected by absorption features. We measured the full width at half-maximum (FWHM) of the emission lines (when they were present) of each spectrum, following a simple Gaussian fitting procedure with the \textit{PySpecKit} Python package \citep{Ginsburg11}. The final classification of every source was done complementing the results of the cross--correlation analysis with visual inspection of every spectrum. To distinguish type 1 and type 2 AGN, we requested that at least one emission line has $\text{FWHM}>1800$ km/s in the rest--frame. In Figure \ref{figure:spec_all_A} we provide the rest--frame optical spectra of our 54 candidates.

\section{Hp--AGN candidates from the RF2 classifier}	\label{RF2_cand}

In Table \ref{table:cand_RF2} we show the list of hp--AGN candidates selected using the RF2 classifier. We provide the equatorial coordinates in degrees (J2000), the $r$ band magnitude, the $g-r$ and $r-i$ optical colors, the most relevant variability features of RF2 ($A_\text{SF}$, $\sigma_{\text{rms}}$, Q31, and $P_{var}$), the predicted class, and the predicted class probability ($P_{\text{RF}}$).

\begin{table*}
\caption{List of hp--AGN candidates selected by the RF2 classifier} \label{table:cand_RF2} 
\begin{center}

\begin{tabular}{lllllllllll} \hline
\hline

RA & DEC & $r$ & $g-r$ & $r-i$ & $A_\text{SF}$ & $\sigma_{\text{rms}}$ & Q31 &$P_{var}$ & class & $P_{\text{RF}}$   \\
\hline

148.314270 & 2.172078 & 20.30 & 0.72 &  0.32 & 0.002 & 4.547e-6 & 0.127 & 0.946 & AGN & 0.86 \\
148.396057 & 0.476811 & 18.22 &  0.41 & 0.14 & 0.107 & 1.323e-6 & 0.105 & 0.999 & AGN & 0.91 \\ 
148.432525 & 1.672532 &  17.62 & 0.37 & 0.15 & 0.069 &  -3.904e-6 & 0.088 & 0.967 & AGN & 0.82 \\
148.497756 & 0.645931 & 19.79 & 0.22 & 0.31 & 0.239 & 1.555e-5 & 0.189 & 1.000 & AGN & 0.99 \\

 \hline
 \hline

\end{tabular}


\end{center}
\textbf{Note.} Only a portion of this table is shown here to demonstrate its form and content. A machine-readable version of the full table is available. 

\end{table*}

\end{document}